\documentclass[fleqn,usenatbib]{mnras}

\usepackage{color}
\usepackage{amssymb}
\usepackage[pdftex]{graphicx}
\usepackage{epstopdf}
\usepackage{amssymb}
\usepackage{aecompl}
\usepackage{soul}
\usepackage{amsmath}
\usepackage{caption}
\usepackage{float}

\usepackage[titletoc]{appendix}

\usepackage{eso-pic}% http://ctan.org/pkg/eso-pic

\AddToShipoutPictureBG*{%
  \AtPageUpperLeft{%
    \hspace{0.75\paperwidth}%
    \raisebox{-1.5\baselineskip}{%
      \makebox[0pt][l]{\textnormal{DES-2021-0641}}
}}}%

\AddToShipoutPictureBG*{%
  \AtPageUpperLeft{%
    \hspace{0.75\paperwidth}%
    \raisebox{-2.5\baselineskip}{%
      \makebox[0pt][l]{\textnormal{FERMILAB-PUB-21-051-AE}}
}}}%

\title[Type Ia SN brightness correlates with host galaxy dust]
{The Dark Energy Survey Supernova Program results: Type Ia Supernova brightness correlates with host galaxy dust}

\author[Meldorf, Palmese, Brout, et al.]{
\parbox{\textwidth}{
\Large
C. Meldorf$^{1,2,3}$, 
A.~Palmese$^{2,3,4}$\thanks{E-mail: \url{palmese@berkeley.edu}}\thanks{NASA Einstein Fellow},
D. Brout$^{5}$\thanks{E-mail: \url{dillon.brout@cfa.harvard.edu}}\footnotemark[2], 
R.~Chen,$^{6}$
D.~Scolnic,$^{6}$
L.~Kelsey,$^{7}$
L.~Galbany,$^{8,9}$
W.~G.~Hartley,$^{10}$
T.~M.~Davis,$^{11}$
A.~Drlica-Wagner,$^{1,2,3}$
M.~Vincenzi,$^{6}$
J.~Annis,$^{2}$
M.~Dixon,$^{12}$
O.~Graur,$^{7}$
C.~Lidman,$^{13,14}$
A.~M\"oller,$^{12}$
P.~Nugent,$^{15}$
B.~Rose,$^{6}$
M.~Smith,$^{16}$
S.~Allam,$^{2}$
D.~L.~Tucker,$^{2}$
J.~Asorey,$^{17}$
J.~Calcino,$^{11}$
D.~Carollo,$^{18}$
K.~Glazebrook,$^{12}$
G.~F.~Lewis,$^{19}$
G.~Taylor,$^{14}$
B.~E.~Tucker,$^{14}$
A.~G.~Kim,$^{15}$
H.~T.~Diehl,$^{2}$
M.~Aguena,$^{20}$
F.~Andrade-Oliveira,$^{21}$
D.~Bacon,$^{7}$
E.~Bertin,$^{22,23}$
S.~Bocquet,$^{24}$
D.~Brooks,$^{25}$
D.~L.~Burke,$^{26,27}$
J.~Carretero,$^{28}$
M.~Carrasco~Kind,$^{29,30}$
F.~J.~Castander,$^{8,9}$
M.~Costanzi,$^{31,18,32}$
L.~N.~da Costa,$^{20}$
S.~Desai,$^{33}$
P.~Doel,$^{25}$
S.~Everett,$^{34}$
I.~Ferrero,$^{35}$
D.~Friedel,$^{29}$
J.~Frieman,$^{2,3}$
J.~Garc\'ia-Bellido,$^{36}$
M.~Gatti,$^{37}$
D.~Gruen,$^{24}$
J.~Gschwend,$^{20,38}$
G.~Gutierrez,$^{2}$
S.~R.~Hinton,$^{11}$
D.~L.~Hollowood,$^{39}$
K.~Honscheid,$^{40,41}$
D.~J.~James,$^{5}$
K.~Kuehn,$^{42,43}$
M.~March,$^{37}$
J.~L.~Marshall,$^{44}$
F.~Menanteau,$^{29,30}$
R.~Miquel,$^{45,28}$
R.~Morgan,$^{46}$
F.~Paz-Chinch\'{o}n,$^{29,47}$
M.~E.~S.~Pereira,$^{48}$
E.~Sanchez,$^{17}$
V.~Scarpine,$^{2}$
I.~Sevilla-Noarbe,$^{17}$
E.~Suchyta,$^{49}$
G.~Tarle,$^{21}$
and T.~N.~Varga$^{50,51,52}$
\begin{center} (DES Collaboration) \end{center}
}\vspace{0.3cm} \\
{\small\emph{(Affiliations are listed at the end of paper)} }}

\begin{document}
\maketitle

\pubyear{2021}

\label{firstpage}

\begin{abstract}
\noindent 
Cosmological analyses with type Ia supernovae (SNe Ia) often assume a single empirical relation between color and luminosity ($\beta$) and do not account for varying host-galaxy dust properties. However, from studies of dust in large samples of galaxies, it is known that dust attenuation can vary significantly. Here we take advantage of state-of-the-art modeling of galaxy properties to characterize dust parameters (dust attenuation $A_V$, and a parameter describing the dust law slope $R_V$) for the Dark Energy Survey (DES) SN Ia host galaxies using the publicly available \texttt{BAGPIPES} code. Utilizing optical and infrared data of the hosts alone, we find three key aspects of host dust that impact SN Ia cosmology: 1) there exists a large range ($\sim1-6$) of host $R_V$ 2) high stellar mass hosts have $R_V$ on average $\sim0.7$ lower than that of low-mass hosts 3) there is a significant ($>3\sigma$) correlation between the Hubble diagram residuals of red SNe Ia that when corrected for reduces scatter by $\sim13\%$ and the significance of the ``mass step'' to $\sim1\sigma$. These represent independent confirmations of recent predictions based on dust that attempted to explain the puzzling ``mass step'' and intrinsic scatter ($\sigma_{\rm int}$) in SN Ia analyses. We also find that red-sequence galaxies have both lower and more peaked dust law slope distributions on average in comparison to non red-sequence galaxies.
 We find that the SN Ia $\beta$ and $\sigma_{\rm int}$ both differ by $>3\sigma$ when determined separately for 
red-sequence galaxy and all other galaxy hosts. The agreement between fitted host-$R_V$ and SN Ia $\beta$ \& $\sigma_{\rm int}$ suggests that host dust properties play a major role in SN Ia color-luminosity standardization and supports the claim that SN Ia intrinsic scatter is driven by $R_V$ variation.

\end{abstract}


\section{Introduction}

Type Ia Supernovae (SNe Ia) have been used for decades as standard candles \citep{riess98,perlmutter99,KEYPAPER,Brout2022PantheonPlus}, and they remain a powerful cosmological tool to probe the expansion of the Universe, along with many other measurements, including those from the Cosmic Microwave Background \citep{planck18}, Baryonic Acoustic Oscillations \citep{Eisenstein_2005}, and more recently, gravitational wave standard sirens (\citealt{2017Natur.551...85A}; \citealt*{darksiren1}; \citealt{2021arXiv211106445P}). 
A deep understanding of the factors that cause variation in a standard candle's luminosity and colour is paramount to increase the level of accuracy and precision of inferred distances and cosmological parameters. 

While photometric calibration has historically been thought of as the dominant source of systematic uncertainty, recent analyses (e.g., \citealt{Brout18-SYS}) have found that empirical modelling of the residual intrinsic brightness variations after standardization (`intrinsic scatter') presents an equally important source of systematic uncertainty. { Note that although this scatter is usually referred to as `intrinsic', it may be caused by effects that are external to the SN itself. It is estimated that because of the current lack of understanding of the physical underpinnings of intrinsic scatter, there exists a significant redshift-dependant systematic uncertainty in the estimated SN Ia distances \citep{brout2020dust} therefore a more complete physical description of SN Ia intrinsic brightness variations is critical for the next generation of SN Ia cosmological analyses.}

{ In this work, we focus on the impact of host galaxy dust and its potential to explain the color--luminosity relation, intrinsic scatter, and the puzzling mass step in SN Ia cosmology.} The way dust can affect the observed SN colours and { cause residual scatter in the Hubble diagram} is as follows. Dust obscures light in the rest frame UV-optical wavelengths and thus affects the brightness and spectral energy distribution of objects observed by optical surveys in two ways. The first is extinction, whereby light from observed objects is obfuscated and reddened by intervening dust particles. Secondly, attenuation includes the effect of extinction, but also accounts for the presence of photons scattering off dust and towards the observer. Attenuation and extinction are characterized by laws that describe the difference in luminosity at a given wavelength between the object observed without attenuation/extinction and the observation we see, i.e. with attenuation/extinction. {Inadequately accounting for the differential luminosity due to dust and potential variance in dust properties causes incorrect inferred luminosities in the SN Ia Hubble diagram.}

 { However, it is known that there can exist a large range of both dust attenuation ($A_V$) and dust laws ($R_V$) across galaxies. For example, the Milky Way, Large Magellanic Cloud, and Small Magellanic Cloud encompass a large range of dust laws ($R_V\sim2-5$) \citep{Schlafly16,Gao2013,Yanchulova17}; {similar variations are seen in} 230,000 galaxies in the local universe \citep{Salim18}. Yet, the currently adopted SN Ia methodology does not allow for this freedom in its use of a single colour--luminosity relation.  This presents two problems in SN Ia cosmology.} 
 
 {  First, because the SN Ia empirical models and relations are mostly agnostic to the underlying physical properties, one needs an additional empirical model to describe the residual scatter in standardized brightness beyond measurement uncertainties. The `G10' intrinsic scatter model \citep{Guy2010} prescribes 70\% of the residual scatter to coherent variation and 30\% to chromatic (wavelength dependent) variation, while the `C11' scatter model \citep{chotard11} prescribes only 25\% to coherent variation and 75\% to chromatic variation. The more recent `SUGAR' model \citep{leget} prescribes intrinsic variations to spectral features.} 
 
 { Second, if dust properties vary in a correlated and systematic manner for different galaxy types, it can manifest as residual correlations in the Hubble diagram. This search for correlations between SN Ia standardized brightness with host galaxy properties has been of large focus in the last decade.} Both global and local environmental properties of SN Ia-host galaxies have been shown to correlate with the distance residuals in the Hubble diagram (e.g., \citealt{Hicken09a,Sullivan2010,Lampeitl2010,gupta11,smith12,Childress_2013,Pan_2013,Johansson,Rigault2013, Rigault_2015, Rose_2019, Rigault_2020, kelsey20}). 
Remarkably, the strongest correlation has often been seen to be with a global galaxy property that is hardly directly related to the SN itself, the total stellar mass (although see \citealt{Rigault_2020} and \citealt{kelsey20} for discussions of the strength of specific star formation rate (sSFR) and rest frame colour respectively). Host galaxy stellar mass has thus been most commonly included as an ad-hoc correction to the standardized brightness in SN Ia cosmology analyses in the form of a step function, hence the name ``mass step.'' 

{ The physical origins of these correlations have largely remained elusive and have involved novel and sometimes complex interpretations. For the `mass step,'} {typically defined} around stellar masses of $M_{\star}\sim 10^{10}M_{\odot}$, several possibilities have been explored in the literature. For example, the age of the progenitor could be related to the age of the overall stellar population of the galaxy, which is often related to its stellar mass \citep{Kelly2010,Childress_2013,Rigault2013, Graur15}. It is also possible that SNe Ia come in different subclasses, for example those that originate from sub-Chandrasekhar mass and Chandrasekhar mass white dwarfs \citep{polin19,polin21}, which show intrinsically different properties. {In addition, recent studies from \citet{Twins1, Twins2} have used manifold learning techniques in an attempt to model intrinsic spectral diversity of SNe Ia by using pairs of similar (dubbed ``twin") supernovae, in a process called Twins Embedding. These studies indicate that models accurately predicting spectra of Type Ia SNe are highly nonlinear in their parameters. Using these models can halve the size of the mass step (0.040 $\pm$ 0.020 mag compared to 0.092 $\pm$ 0.024 mag) but not render it insignificant, hinting that an additional effect potentially unaccounted for in these models, is also driving the presence of the mass step. }

Recently, \citet{brout2020dust} (Hereafter BS21) suggest a physically motivated dust-based model can explain the colour dependent intrinsic scatter as resulting from a wide range of dust laws and the mass step as an artifact of different dust distributions in different mass galaxies. Neither of these concepts are particularly novel. For example, dust parameters have been found to correlate even more strongly with other physical parameters of interest, like the stellar mass. 
\citet{Salim18} find a positive correlation between the stellar mass and the dust column density of galaxies (as given by the parameter $A_V$). In addition, they find lower mass galaxies tend to have steeper attenuation curves (measured by the parameter $R_V$, roughly consistent with the independent prediction by BS21). 

\cite{popovic21} find similar results using SNe Ia with higher certainty using new forward-modeling fitting method and \cite{popovic21bbc} find that when properly accounting for these host-galaxy correlations with SN properties in simulations, there is no longer a need for ad-hoc corrections. While both BS21 and \cite{popovic21} were able to fit the global dust distributions affecting the SNe Ia, they were unable to distinguish whether the dust properties that they have recovered are intrinsic to the SN (i.e. due to the dust produced by the SN explosion itself), the immediate circumstellar environment, or if they are representative of global galactic properties. In addition, Wiseman et al. (\textit{submitted}) show that the $R_V$ varying with the age of the host galaxy is also able to account for the observed SN brightness differences when considering SNe split by both their host stellar mass as well as their host colour. 

A dust-based SN Ia model could also explain three other key aspects of SN Ia cosmology. First, it could explain why host color and sSFR seem to correlate even stronger with SN Ia Hubble residuals than mass (e.g., \citealt{Rigault_2020,kelsey20}), as these parameters trace the dust distributions. Second, it could explain why bluer, un-extinguished, type Ia SNe tend to make better standard candles {\citep{foley_2011,gonzalezgaitan2020effects, Kelsey_2021,Siebert20}.} Lastly it could explain why observations in the NIR are able to achieve less Hubble diagram scatter and smaller `mass steps' (e.g., \citealt{Burns18,johansson2021nearir}), as these observations are less sensitive to dust. 

Though recent studies have attempted to prove a direct link between dust and the observed mass step \citep{Burns18,ponder2020type, Uddin2020, thorp2021testing,johansson2021nearir}, they have not reached definitive consensus as of yet. For example, \cite{ponder2020type, Uddin2020, thorp2021testing} have suggested evidence for the existence of a mass step in the NIR bands which would invalidate dust-based explanations (e.g. \citealt{brout2020dust}) for the step. {However,} \cite{Burns18} and \cite{johansson2021nearir} find reduced or zero evidence for a mass step in the NIR or when accounting properly for dust in the optical. Furthermore, \cite{johansson2021nearir} were unable to reproduce the findings of the contradicting papers \citep{ponder2020type,thorp2021testing, Uddin2020} and note that these papers may be affected by the biased collection of NIR samples largely at high host stellar mass.

{ In this work, we aim to test the BS21 model for the effects of dust on SN Ia observations. Furthermore, we seek to test whether the observational contributions of dust are intrinsic/circumstellar or related to the global galaxy dust as whole.} We derive galaxy parameters for a subsample of the hosts of the Dark Energy Survey (DES; \citealt{descollaboration,2016MNRAS.460.1270D}) 5 Year SN sample where state--of--the--art multi-band photometry is available from deep $ugrizY$ stacks of Dark Energy Camera (DECam; \citealt{flaugher}), and from deep observations from the Visible and Infrared Survey Telescope for Astronomy (VISTA; \citealt{vista}) in the $JHKs$ bands. We note that an analysis of the DES SN hosts from the 3 Year sample has already been carried out in \citet{Smith_2020} (here after S20),\citet{wiseman20,kelsey20}. Our analysis differs from previous works for a number of reasons. First, we pay particular attention to the dust parameters and their correlation with stellar mass, in an attempt to confirm or rule out the arguments from BS21. Moreover, we use a larger number of filters, in particular optical, including $u$, and NIR data, compared to previous works computing stellar masses with DECam bands (\citealt{palmese16,palmese18,wiseman20,kelsey20}; S20), which helps constrain the dust parameters better compared to optical-only estimates (see e.g., Fig 8 of \citealt{Childress_2013}). However this improved dataset is not available over the whole DES SN fields, and we therefore only analyze a subsample (3 out of 10 fields) of the DES SN hosts, totalling 1110 galaxies.

This paper is organized as follows. In \S \ref{sec:data} we describe the DES and VISTA data used. In \S \ref{sec:method} we present the methodology used for the { spectral energy distribution (SED)} fitting and the priors adopted. \S \ref{sec:results} { and \S \ref{sec:DustonMassStep}}  report our results and discussion. We conclude in \S \ref{sec:conc}. In Appendix \ref{sec:appendix}, we discuss differences between different sets of photometric data used in this work. In Appendices \ref{sec:appendixB} and \ref{appc} we include alternate versions of figures presented herein, utilizing different input parameters and { star formation histories (SFHs)}.

\section{Data}\label{sec:data}

\begin{figure}
    \includegraphics[width=0.4\textwidth]{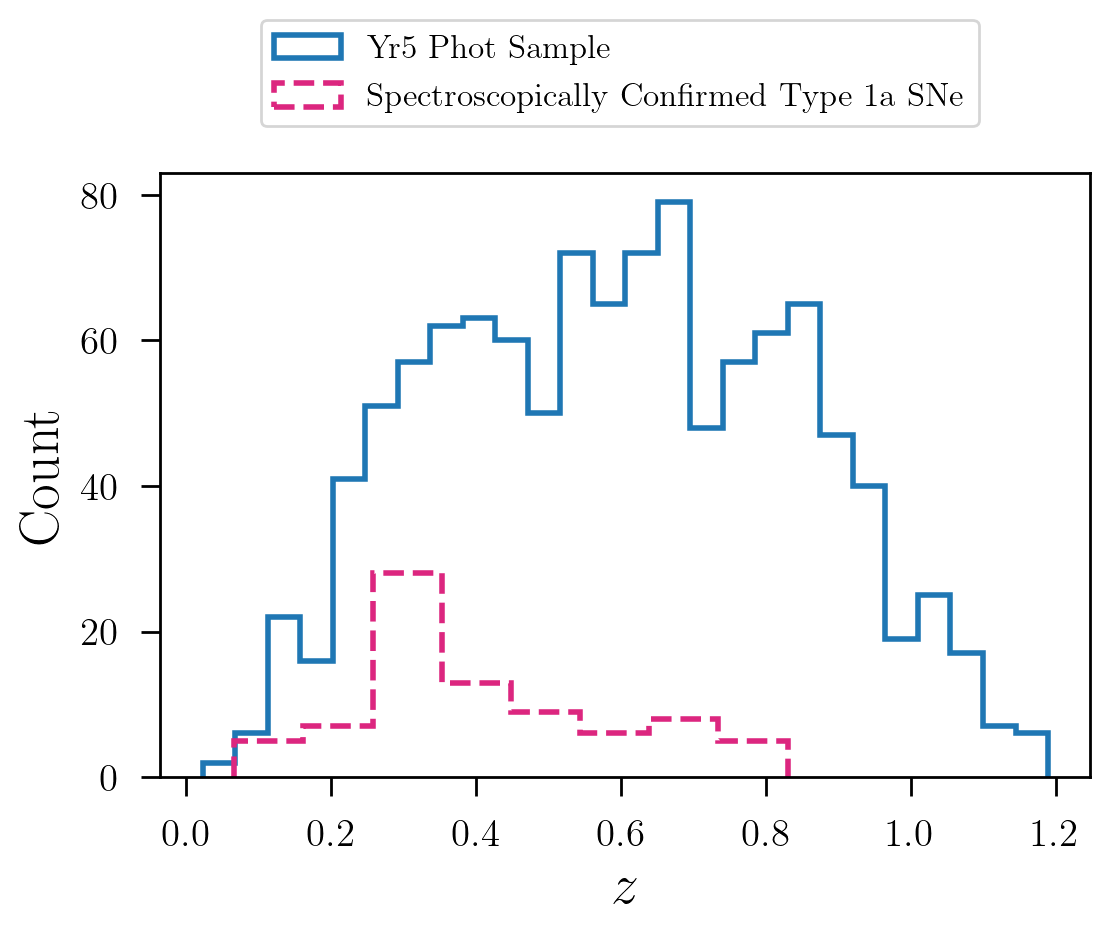}
    \caption{Redshift distribution of the 1110 DES Year 5 ``likely'' SN Ia host galaxies (``{Phot} Sample'', in blue) that overlap with the DES deep fields photometry in \citet*{hartley2020dark}, as well as the distribution of the subsample of 81 galaxies with spectroscopically confirmed Type Ia SNe in the same fields. Redshifts are from OzDES and various public catalogs. }
    \label{redshifts}
\end{figure}

DES (\citealt{descollaboration,2016MNRAS.460.1270D})\footnote{\url{www.darkenergysurvey.org}} is an optical-near-infrared survey that covered 5000 ${\rm deg}^2$ of the South Galactic Cap in $grizY$ bands. DES was carried out using the $\sim3$ $\textrm{deg}^2$ DECam on the Blanco 4-m telescope at the Cerro Tololo Inter-American Observatory.  The DES Supernova Program (DES-SN) has dedicated shallow and deep drilling fields in $griz$ covering 27 ${\rm deg}^2$. SNe are considered as candidates if they satisfy the requirement of two
detections at the same location on two separate nights
in any of the four bands. A subset of all 5 years of DES-SN candidates were selected for spectroscopic follow-up to obtain a type classification and redshift. A detailed overview of the spectroscopic follow-up program as well as a general overview
of the DES-SN program and observing strategy can be
found in S20. In total, $415$ SNe have been spectroscopically confirmed as SN Ia. DES-SN executed its first cosmology analysis \citep{Brout18-SYS} using { scene-modeling} photometry from \cite{Brout18-SMP} for the first 3 years of data (207 spectroscopically confirmed SNe Ia). Here, we utilize difference imaging photometry which is available for all 5 years of SNe Ia and has been shown to be both accurate and achieve similar (within $\sim5$\%) photometric uncertainties (\citealt{Brout18-SMP}) and that the recovered size of the `mass step' is unchanged compared to using { scene modeling} in the DES-SN-3YR analysis \citep{Brout18-SYS}. For the difference imaging sample of all 5 years, 290 SNe pass the same cosmology level quality cuts as \citet{Brout18-SYS}. We analyze the subset of 81 spectroscopically classified SNe that are in fields with VISTA infrared bands' overlap. 

A larger sample of $1110$ host galaxies is also used in this work in order to study the dust properties of typical SN Ia hosts. To select this sample of hosts we follow the simplest version of cuts defined in \citet{Vincenzi2021} for the DES SN photometric sample. In summary, hosts of `likely' SNe Ia are chosen from the associated sample of unclassified transients from the full 5-year DES-SN sample that, when fit with the SALT2 SN Ia light curve model \citep{Guy2007}, are found to be have probability of being a type Ia (\texttt{SNANA FITPROB}) $>0.01$ which is determined from the SALT2 data-model $\chi^2$. Additional requirements are that the transients exhibit SN Ia light curve fit parameters $|c|<0.5$  (c being the colour of the SN) and $|x_1|<5.0$, and fitted SN light-curve peak date uncertainty $< 2$ days. We denote this sample `DES 5 year loose photometric sample', and throughout the text we refer to it as `photometric sample' for simplicity, but note that it differs from the photometrically classified sample in \citet{moller22} and that used for DES cosmology as it does not use photometric classification scores. %, and Chauvenet's criterion. %\citet{Vincenzi2021} find that these requirements alone result in only $\sim5$\% non-Ia SN contamination. 
Because for this larger sample of host galaxies we only study the galaxies themselves and not the SNe candidates, this is sufficient for population studies in this work. Most of the results shown here therefore use this larger sample to maximize completeness and statistics, except for those that rely on SN properties (colour) and SN--derived quantities (the Hubble residuals), where we prioritize the purity of the sample and only select spectroscopically--confirmed SNe Ia. Redshifts of all hosts (also in the larger sample) are obtained from the OzDES program on the Anglo-Australian Telescope (AAT; \citealt{Yuan15,Childress17,lidman2020}) or other external catalogs, as described in more detail in \citet{Vincenzi2021}. The host matching method used is described in \citet{Gupta2016}.

We use the DES Deep Fields { photometric} catalogue \citep{hartley2020dark} prepared for the Year 3 (Y3) weak lensing and clustering cosmology analyses. Briefly, we complement the DES SN-field data in $griz$ with community time DECam data in $u$ band and VISTA data from the Deep Extragalactic Observations (VIDEO; \citealt{video}) and UltraVISTA \citep{McCracken_2012}. The Y3 Deep Fields catalogue consists of 2.8 million sources over 4 fields, covering $\sim 6$ sq deg after masking. One of the fields is in the region of the Cosmic Evolution Survey (COSMOS), which is only used in this analysis for validation tests, while the three of the 10 SN fields covered are C3, E2 and X3 over 4.64 sq. deg. after masking. The depth of the Deep Fields was designed so that the galaxies would have a signal-to-noise $S/N>\sqrt{10}$ times their equivalent in the main survey. Source detection is performed in $riz$ coadded images, having a $10\sigma$ depth of 25.7, 25, and 24.3 mag in $r,i,z$ respectively. The VISTA bands have a 10$\sigma$ depth of 23.0-23.8 ($J$), 22.7-23.3 ($H$) and 22.8-22.9 ($Ks$) in $2''$ apertures, where the ranges take into account the different depths reached in the different fields. Photometry is derived from a bulge plus disk model, fit to the combined $g, r, i$ and $z$-band data using an adapted Multi--Object Fitting (MOF) pipeline, built on the \texttt{ngmix} code.\footnote{\url{https://github.com/esheldon/ngmix}} Measurements in $u$-band and the VISTA bands, $JHKs$, are made by applying this model in a forced-photometry pipeline with the \texttt{fitvd} software.\footnote{\url{https://github.com/esheldon/fitvd}} For a full description of the DES Deep Fields data used here, see \citet*{hartley2020dark}. 
We compare our results that use the Deep Fields photometry from \citet*{hartley2020dark} with those that use the SN coadds in $griz$ from \citet{wiseman20} and the $uJHK$ bands from the Deep Fields. The difference is that the \citet*{hartley2020dark} photometry is drawn from coadd images of the best exposures in each region, and in some cases they are contaminated by the SN light if the best exposures happen to be close in time to the SN, leading to very poor SED fits. The photometry from the \citet{wiseman20} coadds includes only the seasons that do not contain the SN. We find that, despite the difference in the magnitude estimation between the two methods (Source Extractor -- \texttt{SExtractor}, \citealt{SExtractor} -- versus MOF), the recovered galaxy properties for the cases in which the \citet*{hartley2020dark} photometry is not contaminated are consistent within 2$\sigma$ with each other, as we show in Appendix \ref{sec:appendix}. We therefore decide to show results that use the $griz$ \citep{wiseman20} and the $uJHK$ \citep*{hartley2020dark} photometry, in order to exclude the cases of SN contamination. The $uJHK$ photometry comes from observations outside of the nominal DES 5 year (Y5) seasons and therefore they are unlikely to be contaminated by SN light. 

The sample considered in this analysis contains 1110 host galaxies. Most of these are relatively bright galaxies, with $i\lesssim23$, and are therefore bright enough to also be detected in the VISTA bands. About 300 of the galaxies do not have a detection in $JHK$ as they lie in masked regions of the images. We assume an upper limit given by the VISTA depths for these cases, with a large uncertainty of 10 mag. This sample of 1110 galaxies contains a collection of 81 galaxies that hosted spectroscopically confirmed Type Ia SNe. In Figure \ref{redshifts} we compare the distribution of redshifts in the overall sample versus the spectroscopically confirmed subsample.

\begin{figure}
\centering
    \includegraphics[width=0.35\textwidth]{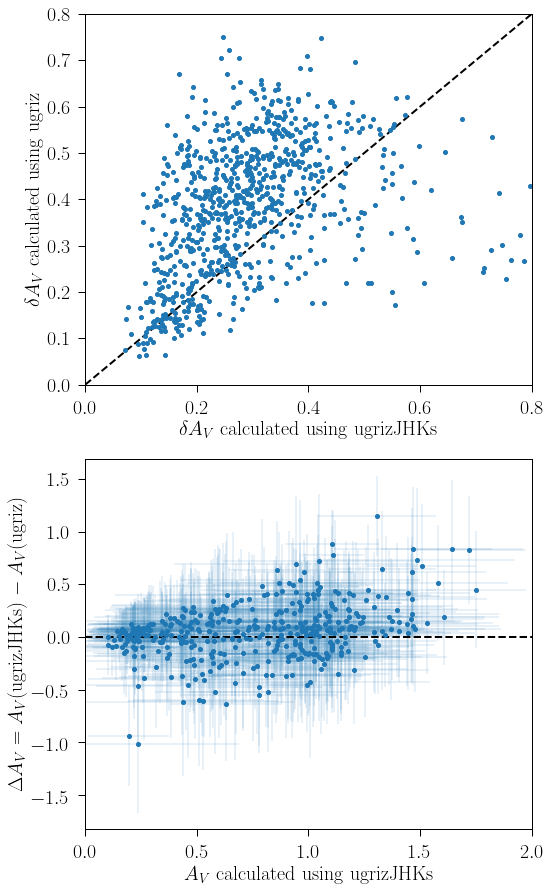}
    \caption{\emph{Top:} Uncertainty in the $A_V$ parameter computed with and without the use of infrared bands. Using the infrared bands consistently leads to lower uncertainties. \emph{Bottom:} residuals between the recovered $A_V$ mean values with and without the use of infrared bands, as a function of the $A_V$ computed using all bands.}
    \label{fig:4}
\end{figure}

{ Derivation of galaxy properties is described in detail in Section \ref{sec:method}, but here we highlight the advantage of combining the VISTA near-infrared photometric data with the DES optical bands.} As interstellar dust tends to redden emitted light, near-infrared data is invaluable to estimate dust parameters. The use of VISTA $J, H,$ and $Ks$ strongly improves the precision of our dust parameter measurements as seen in the top panel of Figure \ref{fig:4}, where uncertainties on $A_V$ are reduced for most galaxies. We note that the inclusion of these bands does not cause significant (at more than $3\sigma$) changes in the mean value of the recovered $A_V$, as is clear from the bottom panel of Figure \ref{fig:4}. 

\begin{table*}\centering
\begin{tabular}{llllll}
\hline
\hline
Parameter & Symbol / Unit & Range & Prior & Prior parameters \\ % %& 
\hline
Redshift & $z$ & (0, 1.5) & Gaussian &  $(\mu, \sigma)=(z_{\rm spec},0.001)$ \\
Mass Formed  & $M_{\rm form}\ /\ \mathrm{M_\odot}$ & (1, $10^{13}$) &Logarithmic    \\
%Stellar Mass  & $M_{\star}\ /\ \mathrm{M_\odot}$ & (1, $10^{13}$) &Logarithmic    \\
Metallicity & $Z\ /\ \mathrm{Z_\odot}$ & (0.1, 2.5) & Logarithmic   \\
$V$-band dust attenuation & $A_V$ / mag & (0, 2) & Uniform   \\
Deviation from Calzetti slope & $\delta$ & ($-1.6$, 0.4) & Uniform     \\
Strength of bump at 2175 $\AA$ & $B$ & 0 & Delta function   \\
\hline
\hline
log-normal SFH max time & $t_{\rm max}$ / Gyr & (0, 15) & Uniform   \\
log-normal SFH FWHM & $\sigma_{\rm SFR}$ / Gyr  & (0, 20) & Uniform    \\
\hline
Exponential SFH age & $t_{age}$ / Gyr & (0, 15) & Uniform     \\
Exponential SFH e-folding time & $\tau$ / Gyr  & (0, 10) & Uniform     \\
\hline
Burst Age & $t_{age}$ / Gyr  & (0.1, 15) & Uniform \\
\hline

\hline
\end{tabular}\caption{Host galaxy parameters estimated using \texttt{BAGPIPES}, to which the DECam+VISTA photometry is input. For each parameter, we provide the prior shape and ranges assumed. The top part of the table shows the parameters we use for all cases. The bottom part shows the parameters needed for the SFHs used in this work. }\label{table:params}
\end{table*}

\section{Method}\label{sec:method}
\subsection{Galaxy modelling}

 In order to derive galaxy properties, we use the Bayesian Analysis of Galaxies for Physical Inference and Parameter EStimation (\texttt{BAGPIPES}; \citealt{Carnall_2018}), a spectral-fitting code used to infer galaxy properties from photometric and spectroscopic data. \texttt{BAGPIPES} employs a fully Bayesian framework and allows us to use sophisticated dust attenuation curves as well as several parametric SFHs. \texttt{BAGPIPES} utilizes the stellar population models derived in \citet{Bruzual_2003} and relies on the \texttt{MultiNest} nested sampling algorithm to obtain the posterior distribution based on model, prior distributions and observational data provided by the user. 
 
 We experiment with a variety of SFHs to determine their effect on the final estimates of the galaxy properties. In particular, we focus on a log-normal SFH. %\apa{- add citations to Carnall 2018, info about the parameter space we use, the dust laws.  Follow https://arxiv.org/pdf/2001.11975.pdf for this part, make a similar table to their tab 1 with the parameter ranges.}
 The log-normal SFH is governed by a peak in star formation rate (henceforth SFR) at some time $t_{\rm max}$, as well as a full width at half maximum parameter, $\sigma_{\rm SFR}$. Both of these parameters are given a uniform prior, over (0, 15) and (0, 20) Gyr, respectively. { These broad priors, which follow other works such as \citet{Carnall_2019}, let us reproduce a wide range of star formation histories, including those that are flat. We extensively test the effect of these priors on the results.} The log-normal SFH is given by the following relation, first presented in \citet{gladders} and revised in \citet{simha2014parametrising}:
\begin{equation}
{\rm SFR}(t) \propto \frac{1}{t} {\rm exp}\Big[ -\frac{({\rm ln}(t)-T_{0,l})^2}{2\tau_l^2} \Big]\, , 
\end{equation}
where {$t$ is the age of the universe,} SFR is the star formation rate, and $\tau_l$ and $T_{0,l}$ are free parameters. We follow \citet{Diemer_2017} and \citet{Carnall_2019} in redefining these parameters in terms of $t_{\rm max}$ and $\sigma_{\rm SFH}$ through the following relations:
\begin{equation}
     t_{\rm max} = e^{T_{0, l}-\tau_{l}^2} \, ,
\end{equation}
and
\begin{equation}
   \sigma_{\rm SFH} = 2t_{\rm max}\hspace{5 pt}{\rm sinh}(\sqrt{2{\rm ln}(2)}\tau_{l}) \, .
\end{equation}
{In addition, we experiment with modifications to the log-normal SFH by adding a delta function ``burst'' in SFR at some time $t_{\rm age}$, defined as the time since the burst. This SFH is referred to as log-normal + Burst.}

The exponential SFH has been widely adopted in the literature (e.g., \citealt{1972A&A....20..383T,1983ApJ...273..105B}; S20). It is parameterized by a start time, $T_{0,e}$, where it increases from a star forming rate of 0 to its maximum value. It then exponentially decays with an $e$-folding time of $\tau$, following: 
 \begin{equation}
\operatorname{SFR}(t) \propto\left\{\begin{array}{cc}
\exp \left(-\frac{t-T_{0}}{\tau}\right) & t>T_{0} \\
0 & t<T_{0} \, .
\end{array}\right.
\end{equation}
\noindent We give $\tau$ a uniform prior of (0, 10) Gyr in general, but we will occasionally use a (0, 2) Gyr uniform prior to compare to other analyses. We give $T_0$ a broad uniform prior over (0, 14) Gyr.

For all SFH models, we assume uniform priors in logarithm space ranging over (1,$10^{13}$) $M_{\odot}$ (i.e., an uninformative prior) and (0.1, 2.5) $Z_{\odot}$ for the total stellar mass formed and for the metallicity $Z$, respectively. The redshift prior is a Gaussian centered at the spectroscopic redshift $z_{\rm spec}$ for each SN, and has a standard deviation of 0.001.

Dust attenuation curves can be parameterized in several ways, and they are often described by the parameters $A_V$ and $R_V$. $A_V$ is the dust attenuation in V band (at $\lambda \simeq 5500 \mathrm{\AA}$). It is often used as an overall normalization constant for the attenuation curve, and is proportional to dust column density. In other words, the more dust between the observer and the observed point, the more dust affects the attenuation curve at all wavelengths. $R_V$ determines the slope of the attenuation curve in the optical wavelengths. $R_V$ is defined as $\frac{A_V}{E(B-V)}$, the ratio between $A_V$ and the reddening $E(B-V)$, which is the difference in extinction in the B and V bands. For our analysis we use the dust attenuation law from \citet{Salim18}, as also done in \citet{carnall20}, governed by the parameters $A_V$, $\delta$ and $B$. The $\delta$ parameter represents the deviation from the Calzetti attenuation curve model \citet{1994ApJ...429..582C} and is directly related to $R_V$ via:
\begin{equation}
    R_{V}=\frac{R_{V, \mathrm{Cal}}}{\left(R_{V, \mathrm{Cal}}+1\right)(4400 / 5500)^{\delta}-R_{V, \mathrm{Cal}}} \, , \label{eq:rv}
\end{equation} 
where $R_{V, \mathrm{Cal}} = 4.05$. It should be noted that this relationship is only valid for $E(B-V) = 1$, and thus should be considered only an approximation for other reddening values.
The $B$ parameter corresponds to the strength of the ``bump" to the total attenuation at 2175 $\mathrm{\AA}$. $B$ is fixed at 0 in our case, so it does not have any effect on the dust law and we do not take it into account in the equations above, nor in the following. This choice is driven from the fact that we are not particularly interested in this parameter, and adding another degree of freedom to the extinction law could be problematic considering the limited number of filters available. We perform a test run leaving $B$ free, and no results are altered enough to be noteworthy. $A_V$ is given a uniform prior over (0,2) while $\delta$ has a uniform prior over (-1.6, 0.4) (following \citealt{salim2020dust}, who revisit the bounds used in \citealt{Salim18}) which corresponds to $R_V$ values in a range (1.2, 7.1). Parameters and priors used in this work are summarized in Table \ref{table:params}. 

The ability of \texttt{BAGPIPES} to recover galaxy properties, including dust, has been extensively studied in \citet{Carnall_2018} and \citet{Carnall_2019} using simulations. \citet{Carnall_2019b} finds trends of galaxy properties with dust attenuation parameters that are in agreement with independent studies. { We also perform a large number of validation tests using COSMOS galaxies, and comparisons with the catalogs of \citet{laigle16,Weaver_2022}, and with \citet{Salim18}. }

\begin{figure*}
    \includegraphics[width=1\textwidth]{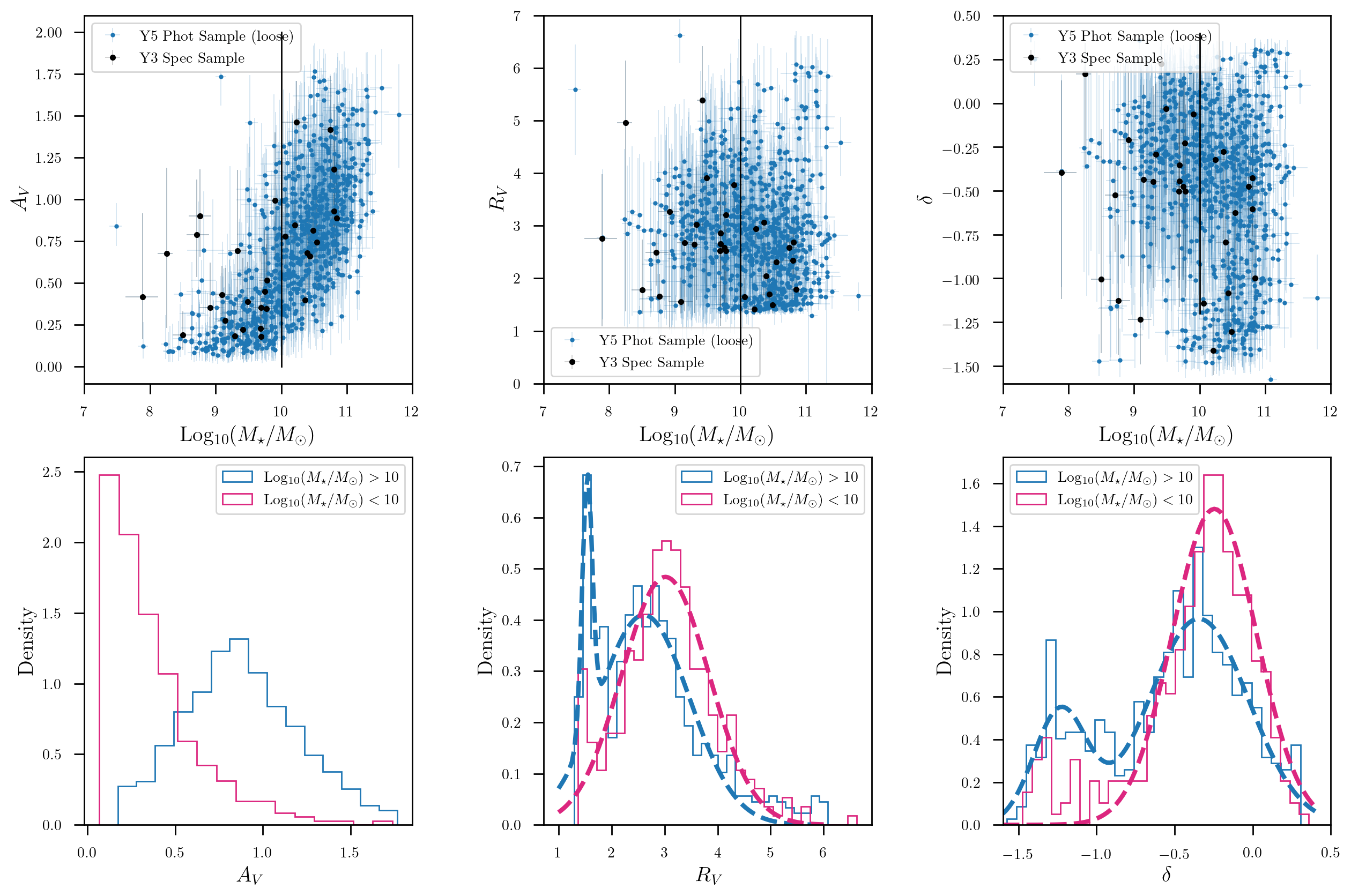}\hspace{0.1in} 
    \caption{\emph{Top:} Dust parameters $A_V$, $R_V$ (as defined in Eq. \ref{eq:rv}), and $\delta$ estimates from this work as a function of stellar mass, derived using a log-normal SFH. Blue points represent galaxies whose photometry data are found in the DES Year 5 sample (Y5 Phot Sample loose), black points are those galaxies found in the DES Year 3 spectroscopic Sample. In all cases we use 3 of the SN fields. { In all six plots, galaxies that did not meet a $\chi^2$ cut are not shown and are not used in calculation of fits.} The vertical line denotes the location of the $10^{10} M_{\odot}$ split. \emph{Bottom:} Distributions of $A_V$,$R_V$, and $\delta$ values above and below the $10^{10} M_{\odot}$ split, the location of the observed mass step. For the $R_V$ and $\delta$ distributions, we include Gaussian curves to highlight the distributions' dependence on mass.}
    \label{fig:5}
\end{figure*}

In this work, we take $\chi^2 < 12.5$ as our standard SED fit cutoff, as this value corresponds to a p-value of $< 0.1$ for 7 degrees of freedom. Out of the 1110 galaxies in the SN loose photometric sample, 870 pass this cut.

\subsection{Supernova Distances}

Our analysis of the spectroscopically-confirmed DES SN Ia light curves (i.e. a subsample of the 1110 SNe that is used in the analyses that rely on Hubble residuals and SN properties) follows that of \cite{Brout18-SYS}. In summary, we fit light curves with the SALT2 model as presented in \cite{Guy2010} and updated in B14 to obtain light curve fit parameters ($m_B$, $c$, $x_1$). Distance moduli ($\mu$) are then inferred with the Tripp estimator \citep{Tripp98} for each SN~Ia by:\
\begin{equation}
\label{eq:tripp}
    \mu = m_B + \alpha_{\rm SALT2} x_1 - \beta_{\rm SALT2} c - M + \delta_{\rm bias}
\end{equation}
where $m_B$ is peak-brightness based off of the light-curve amplitude (log$_{10}(x_0)$) and where $M$ is the absolute magnitude of a SN~Ia with $x_1=c=0$ and noting that $M$ is degenerate with the determination of $H_0$. $\alpha_{\rm SALT2}$ and $\beta_{\rm SALT2}$ are the correlation coefficients that standardize the SNe~Ia and are determined following \cite{Marriner11}, in a similar process to what is done in B14. \cite{Marriner11} minimize a $\chi^2$ expression that depends on the Hubble residuals after applying the Tripp estimator (see Eq.~\ref{eq:tripp}) and normalize residuals by the quadrature sum of the measurement uncertainties and intrinsic scatter. The method separates the sample into redshift bins in order to remove the cosmological dependence of the fitting procedure. The procedure iterates to determine the intrinsic scatter $\sigma_{int}$, $\alpha$, $\beta$ and the resultant distance modulus values.

The observed SN~Ia distances of Eq.~\ref{eq:tripp} are bias corrected by the expectation ($\delta_{\rm bias}$) from simulations of the survey. We employ one crucial difference with the method used in \cite{Brout18-SYS} in accordance with the result found by S20 that the 5-dimensional bias correction {($c,x_1,m_B,\alpha,\beta$)} of \cite{BBC} artificially suppresses the size of the observed correlations between Hubble diagram residuals and host galaxy properties. For this reason we employ a solely redshift-dependent Malmquist bias correction {(1D)} similar to that of \cite{Betoule2014}. While \cite{BBC} found that 1D $\delta_{\rm bias}$ can result in subtle cosmological parameter inaccuracies, we utilize the 1D method because it preserves the observed correlations between light-curve fit parameters ($c$, $x_1$) and host galaxy properties, which is the focus of this paper.

\section{RESULTS I: Host Properties}\label{sec:results}

{ In this Section we show the fitted parameters and resulting properties of the DES host galaxies in the loose photometric sample with VISTA overlap. We examine the relations between dust parameters and mass properties themselves as this has the potential to explain the `mass step'. Finally we explore the relation between host color and the inferred dust parameters.}

\begin{table}
\centering
    \begin{tabular}{|c c c c|} 
    \hline
      Gaussian fit to: & $\mu$ & $\sigma$ & $N$ \\ [0.4ex] 
     \hline\hline
 $R_V$ High $M_\star$ 1 &2.61$\pm$0.07 & 0.86$\pm$0.06 & 0.88$\pm$0.02 \\
 \hline
$R_V$ High $M_\star$ 2 &1.54$\pm$0.02 & 0.10$\pm$0.01 & 0.12$\pm$0.02 \\
\hline
$R_V$ Low $M_\star$  &3.02$\pm$0.05 & 0.82$\pm$0.04 & 1 \\
\hline
$\delta$ High $M_\star$  1 & -0.35$\pm$0.03 & 0.32$\pm$0.03 & 0.77$\pm$0.04 \\
\hline
$\delta$ High $M_\star$  2 & -1.23$\pm$0.04 & 0.17$\pm$0.04 & 0.23$\pm$0.04 \\
\hline
$\delta$ Low $M_\star$  & -0.25$\pm$0.02 & 0.27$\pm$0.01 & 1\\
\hline\hline
    \end{tabular}
    \caption{Gaussian approximation to the $R_V$ and $\delta$ distributions shown in Figure \ref{fig:5} for low and high mass galaxies split at $10^{10} M_{\odot}$. The high mass galaxies are described by two Gaussian distributions. The parameters $\mu$ and $\sigma$ are the mean and standard deviation respectively, while $N$ is a normalization constant ensuring that the total integral of the double Gaussian equal to one.}
\label{tab:GaussFits}
\end{table}

% \begin{figure}
%     \includegraphics[width=0.43\textwidth]{Fig/Fig8_1.6_Av.png}
%     \includegraphics[width=0.43\textwidth]{Fig/Fig8_1.6_delta.png}

%     \caption{Distributions of $A_V$ and $\delta$ for high and low mass galaxies obtained by sampling the posterior distributions of these parameters. The lower and upper bounds of the 68\% HDI are displayed as vertical lines in corresponding colour for low (pink) or high (blue) mass galaxies. The HDI bound values and medians are shown in Table \ref{table:HDI} for both dust parameters and both high and low mass galaxies.} 
%     \label{HDI}

% \end{figure}

\begin{table}
    \begin{tabular}{|c c c c|} 
    \hline
     & Lower bound & Median & Upper bound \\ [0.4ex] 
    \hline\hline
    
    High Mass $A_V$ & 0.41 & 0.88 &1.35 \\
    \hline
    Low Mass $A_V$ & 0.00 &
0.33 &
0.54 \\
    \hline
    % High Mass $R_V$ & 1.008 & 2.547 & 3.409 \\
    % \hline
    % Low Mass $R_V$ &  0.623 & 2.798 & 3.988\\
    High Mass $\delta$ &
   -1.03 &
-0.50 &
0.17 \\
    \hline
    Low Mass $\delta$ &
-0.72 &
-0.34 &
0.40\\
\hline
\end{tabular}

\caption{Lower and upper 68\% HDI values and medians for the dust parameters $A_V$ and $\delta$ and for both high and low mass galaxies.}
\label{table:HDI}
\end{table}

\subsection{The diversity of host galaxy dust parameters}\label{subsec:dustvmass}

{In Figure \ref{fig:5}, we show the recovered distributions of $A_V$, $R_V$, and  $\delta$ values.
We find a large range of $A_V$, $\delta$, and $R_V$ values for each host, specifically for $\delta$ mostly between values of $-1.2$ and 0, or 1 and 4 for $R_V$. This is a wide range in comparison to the distribution found in the Milky Way, $R_V$ between 2.9 and 3.3 \citep{Schlafly16}. Our large range found in $R_V$ agrees with determination of a large $R_V$ spread (1-5) found in \cite{brout2020dust}}.

We note that typical values of $R_V$=3 are often assumed for SN analyses or have found to be tightly peaked in $R_V$ as seen in \cite{mandel2020hierarchical} and \cite{thorp2021testing}. In particular, \cite{thorp2021testing} find that global $R_V$ values lie within the range $R_V = 2.61 \pm 0.21 $, a distribution about ten times narrower than ours. It is however to be noted that the $R_V$ estimation of \cite{thorp2021testing} comes from the SN light curve, while our determination is a global property of the host galaxy. A large spread is also found for a more generic galaxy population such as that of \citet{Salim18}, finding $R_V$ ranging from 1 to 10. Values of $R_V$ that differ from 3.1 have also been measured from the SN data (as opposed to their hosts), e.g., \citet{foley_2011}.

Novel galaxy evolution studies have shown there is a large spread in $R_V$ for different galaxies, and we show, for the first time, that this spread exists in a population of SN hosts in this work. First, deviations from $R_V=3.1$ or 4.05 have been found over the past decades using various galaxy samples (e.g., \citealt{conroy10,salmon,salim16,Salim18}). Moreover, \citet{salim2020dust} find a range of $\delta$ values which is similar to what found here. They show that there exists a correlation between stellar mass and slope of the dust law, as we find. The fact that we reach similar conclusions to what is well known in the galaxy evolution literature motivates the use of these findings to improve SN cosmology analyses, as we show in the following.

\subsection{Dust Parameters and correlation with host mass}\label{subsec:dust}

\subsubsection{Correlation between $R_V$ and host mass}

The correlation between $R_V$, and equivalently $\delta$, and stellar masses is shown in the top middle and right hand side panels of Figure \ref{fig:5}.
Splitting the data at log$_{10}(\frac{M_{\star}}{M_{\odot}}) = 10$, yields two slightly different distributions of $R_V$, where higher mass galaxies have an additional peak around $R_V \approx$ 1.7 compared to the lower mass galaxies, see bottom middle panel of Figure \ref{fig:5}. The possibility that lower mass galaxies may have, on average, higher $R_V$ values and lower $A_V$ values compared to higher mass galaxies, has been noticed in \citet{Salim18} for a generic sample of galaxies. Here, we have confirmed that this is relevant also for a specific selection of SN host galaxies. 

We then quantify the validity of our conclusion that higher and lower mass hosts defined at the mass step correspond to lower and higher $\delta$ (or $R_V$) galaxies, respectively. Utilizing a Kolmogorov–Smirnov (KS) test, we assess the validity of the hypothesis that the $\delta$ (or $R_V$) distribution for lower versus higher mass SN hosts come from the same distribution. {We repeat this test for different values of stellar mass, changing the location where we divide the hosts sample into two.} %, and show the resulting KS statistic and p-values in Figure \ref{fig:10}. 
While the absolute minimum of the p-value occurs at log$_{10}(\frac{M_{\star}}{M_{\odot}})\sim $ 9.7, the distribution for the p-statistic is almost completely flat from log$_{10}(\frac{M_{\star}}{M_{\odot}})\sim$ 9.6 to 10.3. Thus we conclude that the location of the mass step around log$_{10}(\frac{M_{\star}}{M_{\odot}})\sim 10$ corresponds to two different distributions of $R_V$ values. This difference is in line with the expectations from BS21 that higher mass galaxies have lower $R_V$ values, which could explain the mass step.

\subsubsection{Correlation between $A_V$ and host mass}

In the left panels of Figure~\ref{fig:5} we show the relation between $A_V$ and the stellar mass of SN host galaxies, for which we find a strong correlation. $A_V$ values for host galaxies with log$_{10}(\frac{M_{\star}}{M_{\odot}}) \geq $10 are higher on average, and cover a larger range of values, than those below log$_{10}(\frac{M_{\star}}{M_{\odot}}) < $10. This strong correlation is not surprising, and the fact that our findings are similar to that found in previous galaxy evolution studies is a confirmation of the robustness of our galaxy property estimates, and that similar correlations between dust extinction and stellar mass also hold for SN hosts. For example, \citet{Garn_2010,Zahid_2013} find that the main correlation between galaxy properties and dust attenuation in galaxies is that with the stellar mass, as galaxies building up their stellar mass over time, also build up their dust content. The same trends have been recovered by several more recent studies, including \citet{Zahid_2017}. 

We note that the $A_V$ presented here is measured from the host galaxy, not from the SN. Since this is measured as a global parameter using photometry from the entire host galaxy, its value can be different and usually larger than that measured from the SN itself. Our $A_V$ estimate in fact accounts for the dust along the line of sight crossing the entire galaxy, averaged over its light profile, while a generic SN in our sample is more likely to be positioned at some intermediate position along the line of sight within the galaxy, and not close to the nucleus. We find a mean value of $A_V$ of 0.7, with the low mass galaxies being close to 0.4 and the high mass galaxies averaging at 0.9. On the other hand, the typical $A_V$ spread from SNe Ia can be approximated by the term $\beta c$ in the Tripp relation in Eq. (\ref{eq:tripp}). Considering a typical color distribution with standard deviation 0.1 and $\beta\sim 3$ \citep{Scolnic18}, the $\beta c$ term has a standard deviation of $0.3$, which is quite different from the two $A_V$ distributions we find in this work. A comparison with the mean value of $A_V$ from the SNe is not as straightforward as a spread comparison, as the mean is taken out in SALT2 using an empirical relation.

%We provide a fit to the $R_V$ and $\delta$ distributions 
We estimate the location of the peak of the core part of the distributions in $R_V$ and $\delta$ for bins of stellar mass split at  $10^{10} ~M_\odot$, with results reported in Table \ref{tab:GaussFits}. The distributions are binned with Poisson errors. For the low mass bin we use a single Gaussian, while for the high mass bin we assume a double Gaussian, where each Gaussian is normalized with its own normalization factor $N$ summing up to 1 for the overall distribution. Let us first compare the higher $R_V$ and $\delta$ population. We find that the higher mass galaxies present a peak at slightly lower $R_V$ and $\delta$ values (2.6 and $-0.3$, respectively) than their lower mass counterparts (which peak at 3 and $-0.2$). The standard deviation is slightly smaller for the lower mass galaxies (0.8) compared to the higher mass bin (0.9). The lower $R_V$ or $\delta$ population at $R_V\lesssim 2$ or $\delta\lesssim -0.9$ only resembles a Gaussian consistently for any mass split from log$_{10}(\frac{M_{\star}}{M_{\odot}}) = 9.6$ to 10.3 for the higher mass galaxies, which is why we only include this secondary Gaussian for this mass bin. The peak is found to be at $R_V=1.5$ or $\delta=-1.2$, with a standard deviation of 0.1 and 0.2 respectively. The lower $R_V$ and $\delta$ population of galaxies comprises a significant 10-20\% of the high mass galaxies, as shown by the normalization value $N$ in Table \ref{tab:GaussFits}. These results are stable for variation of the mass split location between log$_{10}(\frac{M_{\star}}{M_{\odot}})$ = 9.6 and 10.3. We remind the reader that the $R_V$ distribution findings should be used with care as $R_V$ is estimated through an approximation. The $\delta$ results should be considered instead.

The choice of $\delta$ prior we make in the SED fitting is similar to that in \citet{salim2020dust}, and results in a small pile up of galaxies at the low--end of the prior (see rightmost column in Figure \ref{fig:5}). This effect appears to be somewhat more prominent in the $R_V$ case (see middle column in Figure \ref{fig:5}) compared to $\delta$, but this is expected because $\delta$ is an exponent in Eq. \ref{eq:rv}. Moreover, since our $R_V$ estimation is only an approximation for $E(B-V)=1$, these results are better understood with $\delta$, which we directly estimate. As a test, we extend the prior range on $\delta$ to lower values to approximate $R_V\sim0$. We find that the resulting $\delta$ distribution ends up extending smoothly to lower values, but our conclusions on the $A_V$ and $\delta$ distributions being different for high versus low mass galaxies, and on lower mass galaxies having smaller $A_V$ and larger $\delta$ than higher mass galaxies, are largely unchanged.

{It it important to quantify the impact of uncertainties in the dust parameters estimates in order to understand whether the difference between high and low mass $A_V$ and $\delta$ (or equivalently, $R_V$) distributions is significant. The average of $\sigma_{A_V}$ is $0.3$, meaning that galaxies at the peak of the low mass $A_V$ distribution are $\sim3 \sigma$ away from the peak of the high mass distribution. The average of $\sigma_\delta$ is $0.4$, and $\sim 1$ for $\sigma_{R_V}$. This implies that galaxies with a $\delta$ (similarly for $R_V$) estimate around the peak of the low mass galaxies distribution at $\delta=-0.2$ are statistically different at $\sim 2.5\sigma$ from those in the secondary Gaussian of the high mass sample around $\delta=-1.2$. We therefore expect differences in the distributions to be significant despite the large measurement uncertainties. A test to verify this consists in using the full posterior distributions on the dust parameters.} In Table \ref{table:HDI} we show the 68\% Highest Density Interval (HDI) for the distributions of $A_V$ and $\delta$ that take into account the measurement uncertainties. We construct these distributions by taking 500 samples from the output posteriors of the dust parameters for each galaxy. We provide distributions for $A_V$ and $\delta$ using this method for both high and low mass galaxies in Table \ref{table:HDI}. 
% The lower and upper bounds of the 68\% HDI are displayed as vertical lines for both low and high mass galaxies. 
Applying a KS test to both of these distributions yields p-values much less than 0.05, indicating it is highly likely these two samples were drawn from two different fundamental distributions. \cite{thorp2021testing} find that their fitted $R_V$ is consistent between low-mass and high-mass, contrary to our findings and the findings of \cite{johansson2021nearir}. 
{We note that when measurement uncertainties are taken into account, the 68\% HDI for higher mass galaxies occupies  lower values in the $\delta$ parameter space compared to lower mass galaxies, similarly to the case of $\delta$ point estimates. This supports the arguments of BS21. When it comes to the spread of the distributions, however, we are not able to argue if one of the two populations has a larger scatter.
In other words, with the current precision on dust parameter estimates and host statistics, we are unable to quantify how much of the spread in the distributions is given by intrinsic variations in dust parameters rather than measurement uncertainties.}
%While we note that the main difference in the $R_V$ distributions is given by the different spread, we cannot exclude that this is purely due to higher mass galaxies being on average more luminous, and therefore producing more precise dust parameter constraints compared to the lower mass, less luminous sample. 
In Table \ref{table:HDI}, we report the specific boundary values of the highest density intervals, as well as the median of the distributions.

\section{RESULTS II: Correlations Between Host Properties and SN Ia Properties} \label{sec:DustonMassStep}
{Here we examine the relation between host properties and the SN Ia properties. { First, we explore} the relation between fitted $R_V$ of the host and fitted $\beta_{\rm SALT2}$ from the SNe Ia on similar populations of galaxies. Then, we show correlations between host properties and the Hubble residuals from the SNe in the DES5YR spec sample. Lastly, we show the impact of dust based corrections to the SN Ia standardized luminosity and its effect on the inferred mass step.}

\subsection{Host Dust Drives the SN Ia Color-Luminosity Relation}\label{sec:dustdrives}

\begin{figure}
    \centering
    \includegraphics[width=0.96\linewidth]{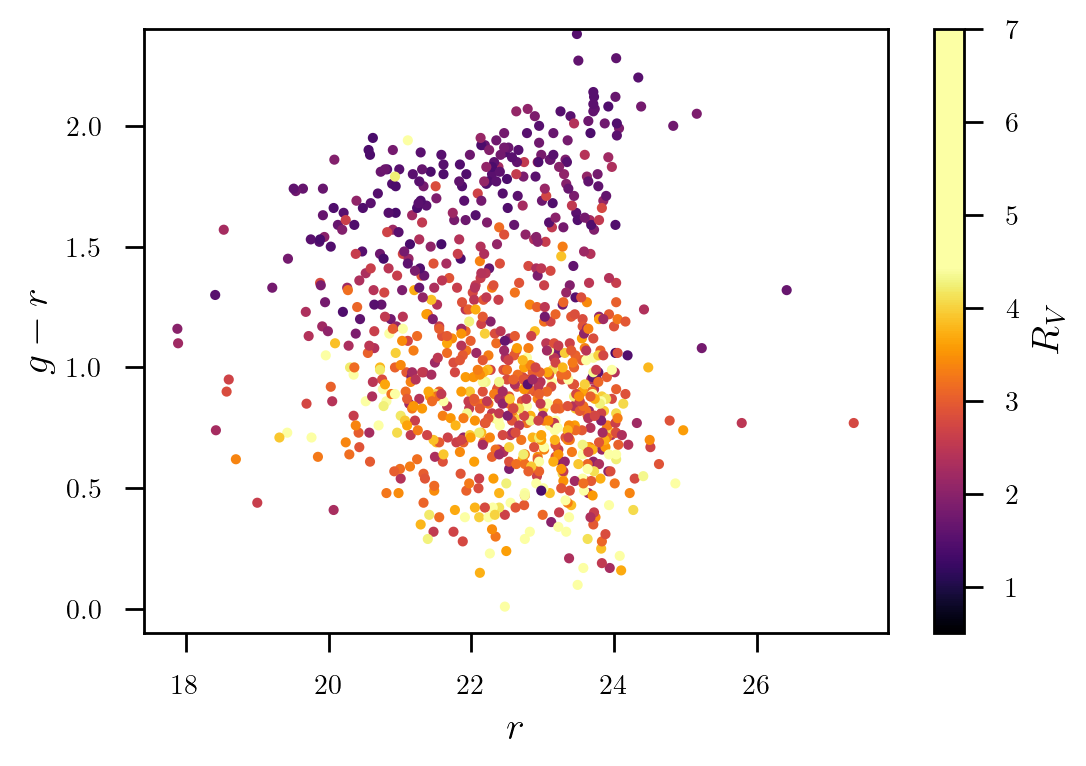}
    \caption{Color-magnitude plot of the hosts considered in this work, color-coded by their best-fit dust law $R_V$ value. The red sequence galaxies (such as the redMaGiC galaxies) clearly represent the largest fraction of low $\delta$ (or similarly, $R_V$) galaxies, meaning that they tend to have a steeper dust law slope.}
    \label{fig:rvhostcolor}
\end{figure}

{ In order to explore the cause of the low $R_V$ values seen in Figure \ref{fig:5}, the recovered host $R_V$ values are plotted as a function of host color. In Figure \ref{fig:rvhostcolor}, it can been seen that there is a dependence of $R_V$ with a split occurring roughly at host color of $g-r = 1.25$. For host colors redder than this split the $\langle R_V \rangle$ = 1.8 and bluer than the split $\langle R_V \rangle$ = 3.2 (shown in Table \ref{tab:rvsplit}). This follows the expectation that steeper dust law slope produces redder SEDs, and that red sequence galaxies are those with a lower $\delta$ or $R_V$. On the other hand, the very high $\delta$ or $R_V$ galaxies are very blue, star forming galaxies.}

{For the subset of DES spectroscopically confirmed SNe Ia in the entire DES Y5 spectroscopic sample (i.e. not only those with VISTA bands analyzed in this work, since here we are not using a galaxy SED fit) of red (81 SNe) and blue (209 SNe) hosts, we perform two separate fits of $\beta_{SALT2}$ when splitting on host $g-r$ color = 1.25. For the SNe Ia in red hosts we find $\beta_{SALT2}=2.09\pm0.15$ and for the SNe Ia in other hosts we find $\beta_{SALT2}=2.71\pm0.11$. This 3.3 sigma difference in $\beta_{SALT2}$ is consistent with the difference found in dust $R_V$ for the same split on host color.} {  We note that $\beta_{\rm SALT2}$ is defined for $B$ band magnitudes and that $\beta_{\rm SALT2} \sim R_B = R_V + 1$.}

\begin{figure}
    \centering
    \includegraphics[width=0.9\linewidth]{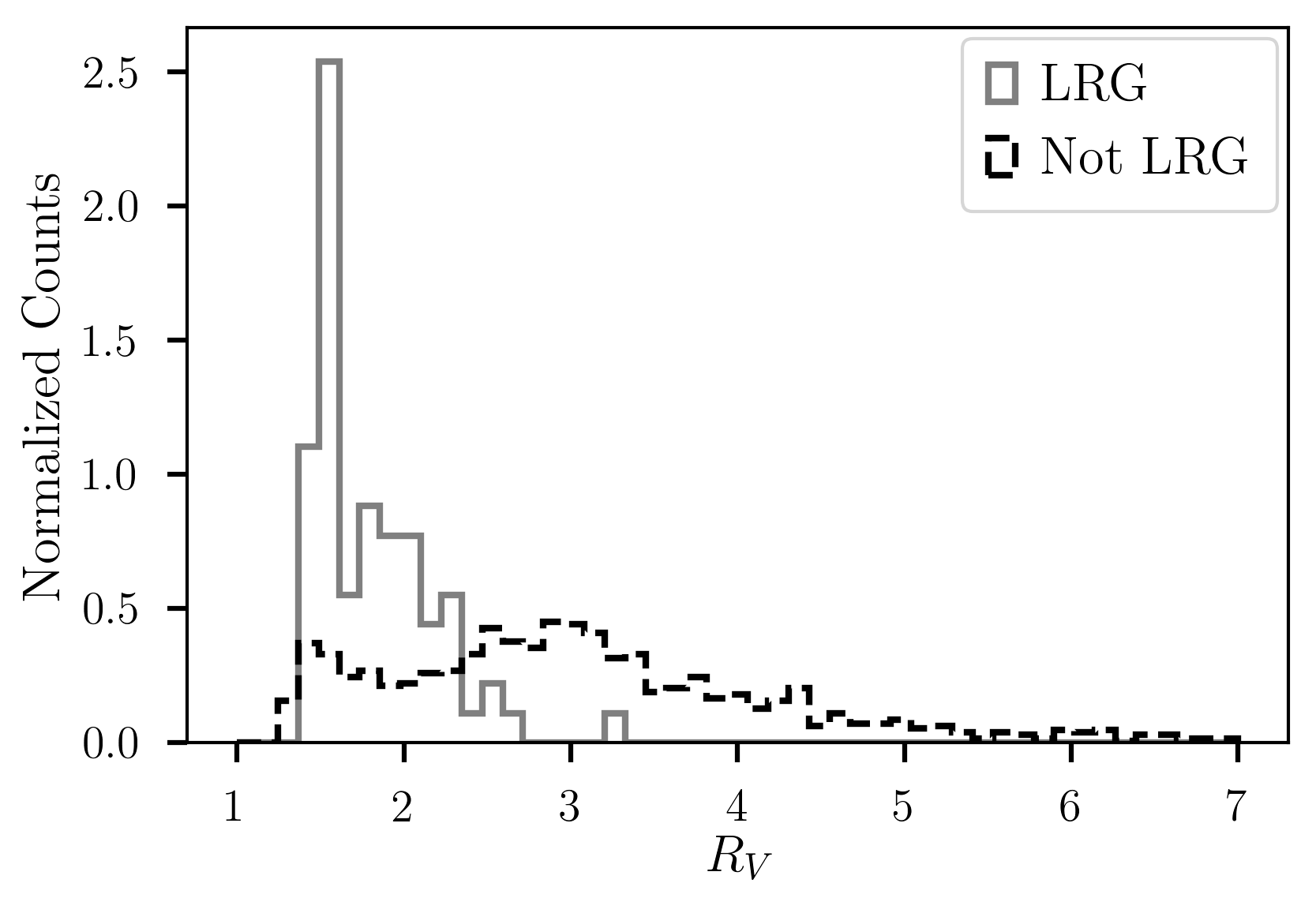}
    \caption{Fitted $R_V$ for galaxies that are matched to the DES redMaGiC catalog (solid grey) and those that are not (dashed black). These galaxies are LRGs selected using the redMaGiC algorithm.}
    \label{fig:rm}
\end{figure}

\begin{table}
    %\centering
    \resizebox{1.05\columnwidth}{!}{%
    \begin{tabular}{| c c c c c |} 
    \hline
     Host Color Split & Host $\langle R_V \rangle$ & Host $\sigma_{R_V}$ & SN Ia $\beta_{\rm SALT2}$ & SN Ia $\sigma_{\rm int}$\\ 
    \hline\hline
    
    $g-r >$ 1.25 & 1.82 & 0.47 & $2.09\pm0.15$& $0.05\pm0.015$ \\
    $g-r <$ 1.25 & 3.15 & 0.90 & $2.71\pm0.11$& $0.10\pm0.007$ \\
    
    \hline
    \end{tabular}
    }
\caption{Fitted distributions of $R_V$ when split on galaxy $g-i$ color. Mean $R_V$ and 1$\sigma$ width $\sigma_{R_V}$ are reported.}
\label{tab:rvsplit}
\end{table}

\begin{figure*}
    %\centering
    \includegraphics[width=1\linewidth]{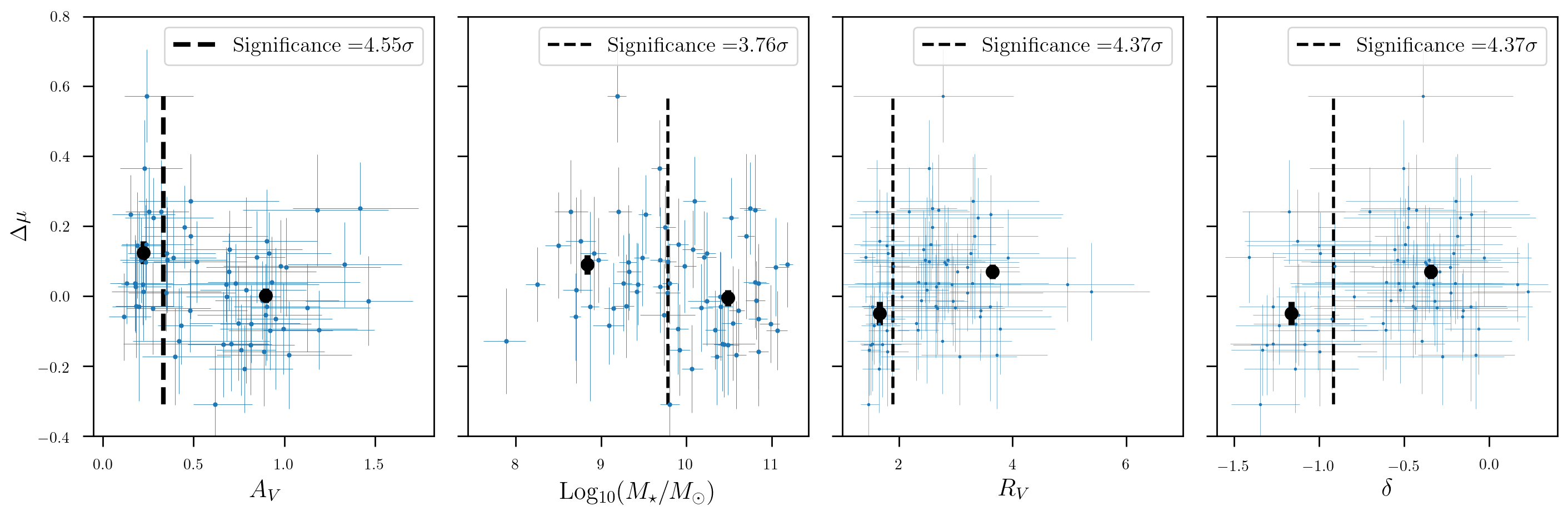}
    \includegraphics[width=1\linewidth]{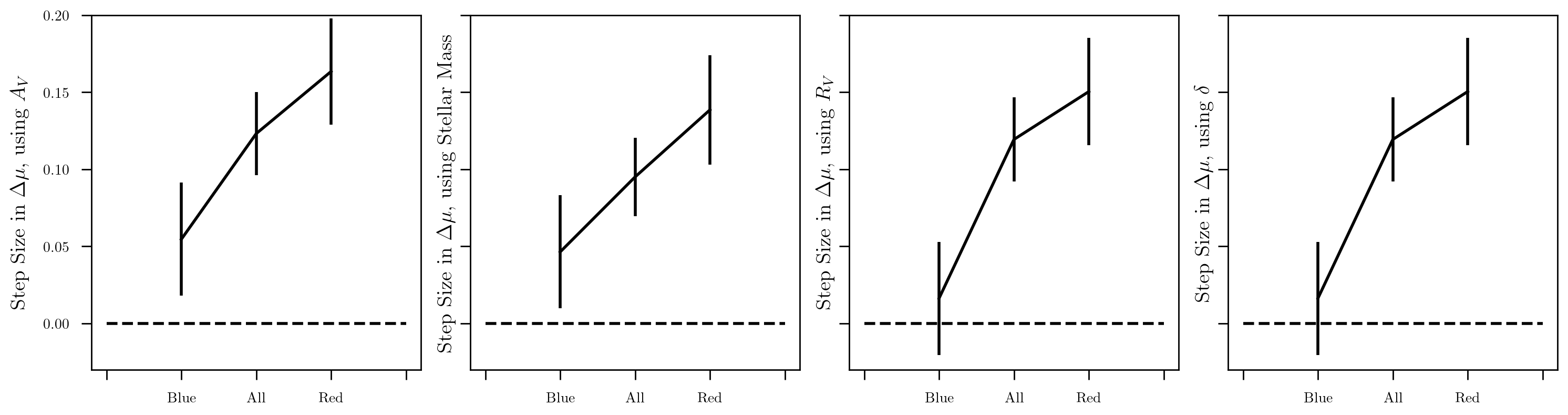}
    \caption{\emph{Top:} Step-like behaviour in Hubble residuals as a function of $A_V$, stellar mass, $R_V$ and $\delta$, calculated using a log-normal SFH. %These are calculated using five SFHs, four of which being exponential with $0<\tau<2$ or $0<\tau<10$ Gyr with $R_V$ free or fixed, the final being log-normal with $R_V$ left free. 
    We scan over each parameter to search for a significant step in the Hubble Residuals, i.e. a certain point where the mean of the Hubble residuals on either side of said point is significantly different. 
    The most significant step found is displayed in each panel, along with its 1$\sigma$ uncertainty on the mean in black. {The location of the splitting that gives the most significant step is displayed as a vertical dashed line.} 
    \emph{Bottom:} Step size in the Hubble Residuals as we move from what we expect to be a less dusty to a more dusty sample. The ``red'' and ``blue'' samples refer to red and blue SNe, having a colour { parameter above or below $c = -0.025$ , respectively.}. The ``All'' sample refers to what was shown in the top row of this figure. 
    % As in Figure \ref{fig:12}, the steps are coloured according to their significance where significance was determined by the step calculated from the SNe and host properties in each subsample. 
    Note that in each case the blue (presumably less dusty) sample yields a step which does not appear to be significant, while the red (presumably more dusty) sample always yields a significant step. We define as significant a step size that it at least 3$\sigma$ away from 0. The errorbars in the plots are 1$\sigma$. A version of this plot with more examples using different SFHs is available in Appendix \ref{appc}.}
    \label{fig:12}
\end{figure*}

We perform the same test using the subsample of the 870 host galaxies that is identified as luminous red galaxies (LRG) by the \texttt{redMaGiC} \citep{redmagic} algorithm applied to the DES wide field survey. \texttt{redMaGiC} identifies red galaxies based on the redMaPPer \citep{Rykoff14} red sequence template, given a minimum luminosity cut. Figure \ref{fig:rm} shows the $R_V$ distribution for the 79 LRG galaxies in our sample (solid line) and for all other galaxies (dashed). From Figure \ref{fig:rm} it is evident that the LRG sample is found to typically have much lower values of $R_V$. For the set of SNe in LRGs,  
\cite{Chenetal} find a very low value of SALT2 $\beta$ $\sim2$ instead of the typical $\beta$ $\sim3$ that has found in past cosmological analyses of SNe Ia from unselected samples (e.g., \citealt{Brout18-SYS}).

These tests mark a concrete advancement over BS21, which was not able to discriminate if the dust that causes a diversity in reddening behavior was due to circumstellar or galactic dust; by connecting the recovered low value of $\beta$ with a predicted low value of $R_V$, this finding points towards galactic dust as a driving force of SN reddening.

\subsection{Host Dust Drives Residual SN Ia Scatter}

{ From the same test of splitting on host color (at $g-r=1.25$) described in Section \ref{sec:dustdrives}, this time we consider the spread in $R_V$ above and below the split and its relation to SN Ia intrinsic scatter in distance. For host colors redder than this split, the 1-$\sigma$ width of the $R_V$ distribution ($\sigma_{R_V}$) is 0.47 and for hosts bluer than the split $\sigma_{R_V}$ is 0.90, roughly twice as wide (Shown in Table \ref{tab:rvsplit}). 

The scatter in the SN Ia standardized distances under the same host galaxy conditions is shown in Table \ref{tab:rvsplit}. We find that for the SN Ia in red hosts the intrinsic scatter ($\sigma_{\rm int}$) is $0.05\pm0.015$ mag whereas for the blue hosts we find the SN Ia $\sigma_{\rm int}$ is $0.10\pm0.007$ mag. These values are distinctly different, at 3$\sigma$ significance. The scatter of the SNe Ia is twice as small for the sample in red hosts in comparison to blue hosts, which is consistent with the scatter found for $R_V$, suggesting that dust is indeed driving the scatter in the SN Ia luminosities.}

\subsection{Host Dust Corrections for Current SN Ia Analyses} \label{subsec:DustCorrection}

\begin{figure*}
    \centering
    \includegraphics[width=1\linewidth]{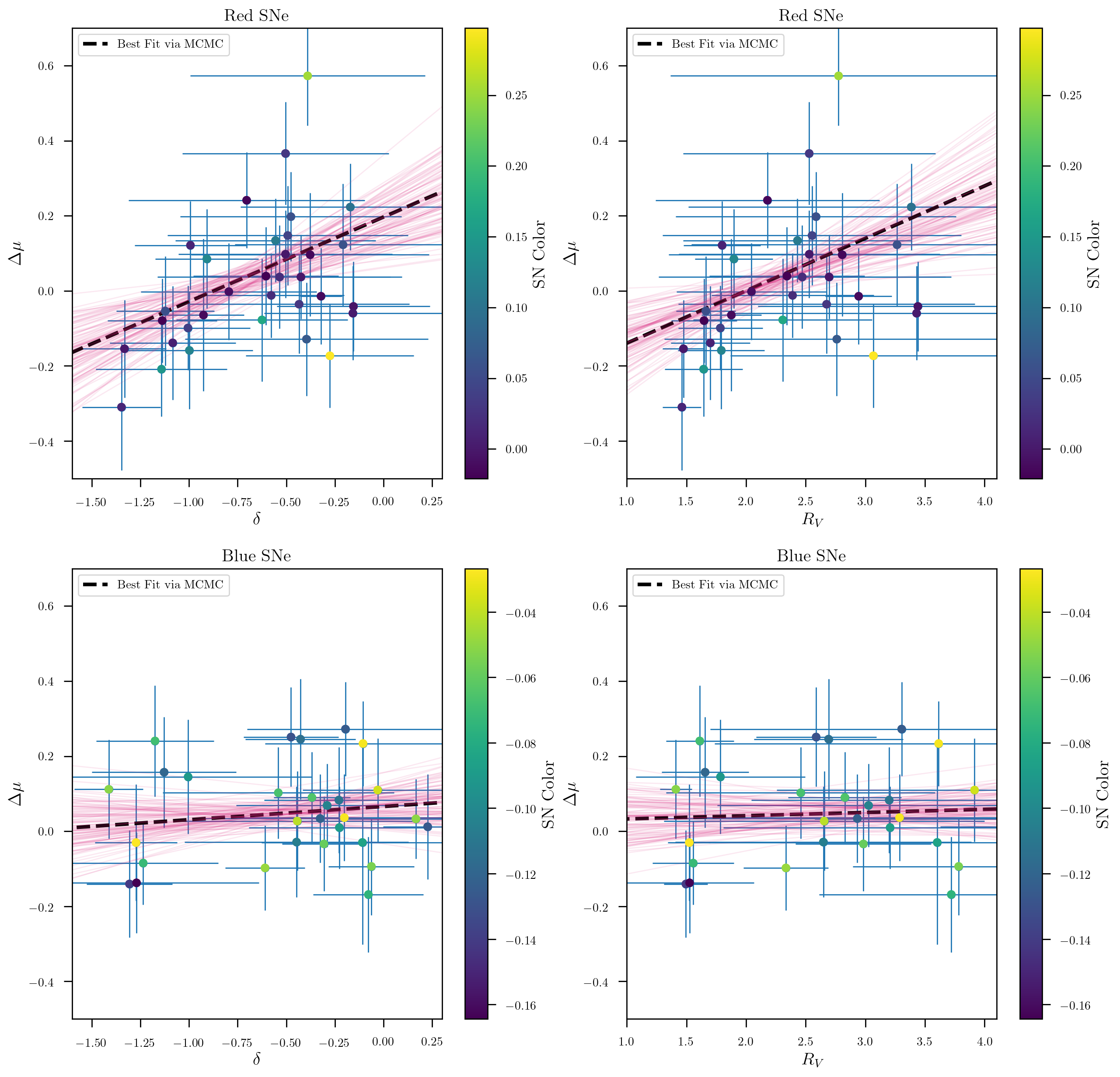}
    \caption{\emph{Top:} Correlation between the host galaxy dust parameters $\delta$ (left) and $R_V$ (right) and Hubble Residuals ($\Delta\mu$) for red (and presumably reddened by dust) SNe with SALT2 $c>-0.025$, with the best fit linear relation (dashed line). The thinner lines represent an MCMC sampling of the posteriors for the slope and intercept of the relation. {The slope is ${m} = 0.219_{-0.069}^{+0.076}$
    for the relation with $\delta$ (which is $\gtrsim 3\sigma$ away from a flat relation) while the $R_V$ relation has a slope of ${m} = 0.122_{-0.046}^{+0.056}
$}. The colour of the data points represents the colour of the SN.
    \emph{Bottom:} same as the top panels, but for the blue SNe with SALT2 $c<-0.025$, which are less likely to be affected by dust than their redder counterparts. The trend between $\delta$ or $R_V$ and $\Delta\mu$ vanishes, giving random scatter around zero with  ${m} = 0.008_{-0.024}^{+0.024}$ for the $R_V$ case and ${m} = 0.033_{-0.051}^{+0.052}$ for the $\delta$ case.}
    \label{fig:RvDeltaMu}
\end{figure*}

In this subsection we seek to identify correlations between various host galaxy { dust parameters and the SN Ia Hubble Residuals ($\Delta\mu$) to be made as ad-hoc corrections in current analysis methodologies.} 
We note that the residuals here are only computed for the subsample of 81 SN Ia candidates from DES Year 5 that have been spectroscopically classified and that fall within the DES deep fields with available VISTA data. 

In the top panels of Figure \ref{fig:12} we analyze the step--like properties in the Hubble Residuals with respect to several host-galaxy parameters. We scan over the range of each parameter, finding the difference in the mean on either side of a step, and reporting results for the step position that maximizes significance. All steps considered here (i.e. stellar mass, $A_V$ and $\delta$ - which is equivalent to $R_V$) are found to be significant and are plotted in black in the top row of Figure \ref{fig:12}. 
We note that the step in dust parameters is as significant as the step in stellar mass, and even more significant in the case of $\delta$ or $R_V$.

In order to further understand the role of dust, we divide the SNe into red and blue SNe, by splitting them in colour at $c = -0.025$. \cite{brout2020dust} determined that this value separates the SNe that are significantly affected by dust (red sample) from those that are closer to their intrinsic colour (blue sample). We decided to adopt this split rather than trying to model the entire sample as a whole for simplicity: we do not expect the blue SN sample to be significantly sensitive to the dust properties and so we expect it to mostly add noise when measuring relations between Hubble residuals and dust parameters. We report the steps in host properties for blue and red SNe separately in the bottom row of Figure \ref{fig:12}. For all galaxy parameters, the step in Hubble residuals for the blue SN sample is not significant ($<2\sigma$) while it is significant for the red SN sample, further suggesting that it is the SNe extinguished by dust that drive the observed SN Ia relations with host properties. Dixon et al (in prep) use host galaxy spectra instead of photometry to derive galaxy parameters and find similar trends between Hubble residuals and host galaxy dust, which when split by SN Ia color, are only present in the redder SNe Ia.

%This process was repeated for parameters computed with several different SFHs: an exponential with $0<\tau<2$ Gyr with $R_V$ both free and fixed, an exponential with $0<\tau<10$ Gyr with $R_V$ free and fixed, and a log-normal SFH with $R_V$ left free. The form of the SFH has a noticeable effect on the significance of the steps in the Hubble Residuals. We find a significant step with respect to mass in every SFH, but for those where $\tau$ is restricted to [0,2] Gyr there is no significant step in the dust parameters. In every other case where there is no such restriction on $\tau$ there is a significant step in all dust parameters. This is important to note for the DES cosmology, as the $0 < \tau < 2$ Gyr prior is the one used in S20, where the mass step is described in detail as resulting from the DES 3Yr data.

In addition, we specifically analyze a linear correlation between the dust law slope parameter $\delta$ (and $R_V$) with Hubble Residuals ($\Delta\mu$). When performing a cut on SNe based on colour at $c = -0.025$ as done above, a significant linear correlation between $\delta$ or $R_V$ and $\Delta\mu$ becomes apparent (Top of Figure \ref{fig:RvDeltaMu}). For the $\delta$ relation with the red SNe Ia ($c > -0.025$), the slope and 1$\sigma$ confidence interval is ${m} = 0.219_{-0.069}^{+0.076}$, and the intercept is ${b} = 0.179_{-0.053}^{+0.058}$, using an MCMC linear fitting algorithm. The maximum likelihood fit is $\sim 3\sigma$ away from a relation with $m=0$, and we therefore deem the correlation significant. For blue SNe ($c < -0.025$) we find that the correlation is absent in this sample: ${m} = 0.033_{-0.051}^{+0.052}, \,b= 0.063_{-0.035}^{+0.036}$, as can be seen from the bottom panel of Figure \ref{fig:RvDeltaMu}.  We stress that the correlation found for the red SNe is not a result of outliers skewing the overall slope. Repeating the fit with outliers removed (any galaxy that was an outlier in $\delta, \Delta\mu,$ or colour was removed, { and one 3$\sigma$ SN Ia $\Delta\mu$ outlier was removed}) achieved consistent results. This result is compelling as it had not been detected in previous works: for example, \citet{Pan_2013} find no correlation between SN colour and reddening, and conclude that extinction is not a main cause of the diversity of SN properties and of the mass step.

We remind the reader that the $R_V$ or $\delta$ we compute here reflect a global quantity from the galaxy, and will not fully describe the change in SN colour compared to its intrinsic value: there are other factors in play, including the dust at the location of the SN, and Circumstellar Material (CSM) interactions. On the other hand, it is worth stressing that our dust measurements from the host galaxy do not use any information about the SN, and are still found to correlate with the SN Hubble residuals.

\begin{figure*}
    \centering
    \includegraphics[width=.9\linewidth]{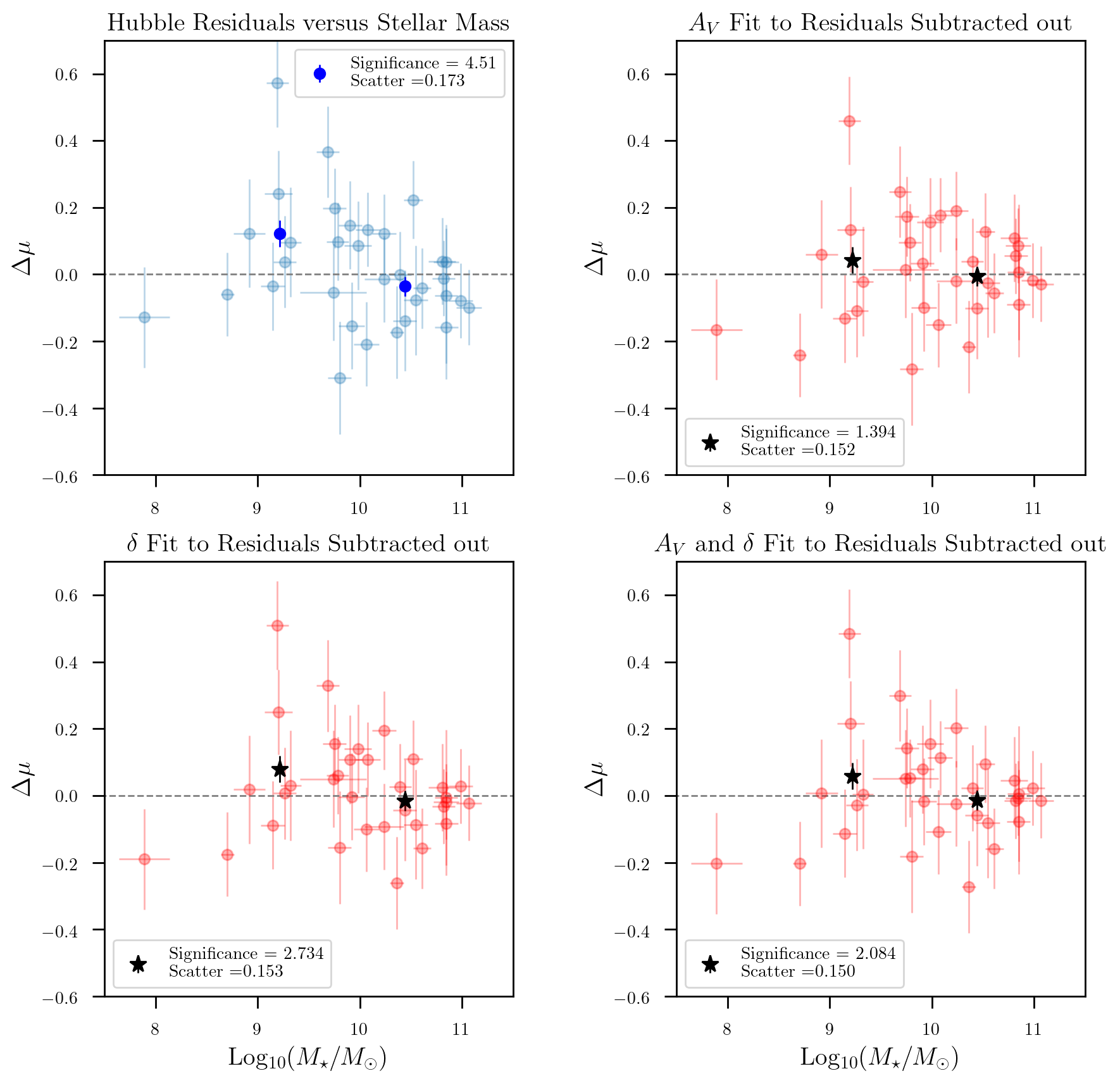}
    \caption{Mass step for red SNe Ia in our sample (color $c>-0.025$), with corrections that use host galaxy dust information. \emph{Top left:} unaltered Hubble Residuals as a function of stellar mass for red SNe. The largest step occurs at log$_{10}(M_{\star}/M_{\odot}) = 9.8$, with a significance of 4.5$\sigma$. The darker blue points show the mean of the data on each side of the split.  
    \emph{Top right:} we fit a 2nd-order polynomial to the Hubble Residuals as a function of $A_V$, and subtract this relationship out for each individual galaxy based on its measured $A_V$. This change reduces the significance of the mass step down to 1.4$\sigma$, as it can be seen from the black star markers, showing the new mean values on either side of the split, while also decreasing the scatter to 0.15. 
    \emph{Bottom left:} another correction to the mass step, this time subtracting out the linear relation fit between Hubble residuals and $\delta$.
    This correction also reduces step size and scatter, down to $2.7\sigma$ and 0.15 respectively.
    \emph{Bottom right:} we fit the Hubble residuals as a function of both $A_V$ and $\delta$, and correct the mass step plot with its best fit. This correction provides the most interesting results, with the step being reduced to 2$\sigma$ and the scatter being reduced to 0.15. We claim that the significant scatter reduction coupled with mass step decrease points to an interesting connection between the mass step and dust parameters, and to promising applications of host galaxy dust estimates to reduce the scatter in the Hubble diagram. }
    \label{dustfits}
\end{figure*}

\subsection{Impact of Host Dust Corrections on the `Mass Step'} \label{subsec:DustonMassStep}

Finally, we apply { fits to} the observed correlations between the dust parameters and the Hubble residuals as corrections { to the SN Ia luminosities}, to determine if accounting for host-galaxy dust can explain the mass step. 
In the top left panel of Figure \ref{dustfits}, we { show uncorrected} Hubble Residuals of 33 red DES SNe Ia (taken from the total 81 DES SNe Ia after their host galaxy fits are subjected to a $\chi^2$ cut and a cut on color) versus the stellar mass of their host galaxies. By scanning from low to high mass, we determine that the highest significance mass step occurs at log$_{10}(M_{\star}/M_{\odot})=9.8$ with a significance of 3.76$\sigma$. The RMS scatter in the data is 0.17. The darker blue points represent the mean $\Delta \mu$ value on either side of the mass step split.

After fitting the $\Delta\mu$-$\delta$ relation as described in Section \label{subsec:DustCorrection}, we fit a second order polynomial to the $\Delta\mu$-$A_V$ data that are shown in Fig. \ref{fig:12}. We then subtract the recovered relations out of the $\Delta\mu$'s based on each galaxy's $A_V$ and refit the mass step. We choose a second order polynomial instead of a linear relation for $A_V$ because the data shows a parabolic shape, and because the Akaike Information Criterion (AIC) is improved compared to the linear regression. The best fit relation is: $\Delta\mu=0.333A_V^2-0.657A_V+0.253$. Subtracting out the $\Delta\mu$ value based on the best-fit relation and its $A_V$ value results in the red points in the top-right plot of Figure \ref{dustfits}. Remarkably, this procedure renders the mass step much less significant: the difference between the means on either side of the mass step split { after the $A_V$-$\Delta\mu$ correction} drops to a significance of only 1.4 $\sigma$ {(black points)}. Furthermore, the scatter decreases from 0.173 mag to 0.152 mag, meaning that the decrease in the mass step { significance} is not due to an increase in scatter. The bottom left panel of Figure \ref{dustfits} shows the results of subtracting out a linear $\delta$ correlation shown in Figure \ref{fig:RvDeltaMu}. Again, both mass-step size and RMS scatter in $\Delta\mu$ are reduced to 2.7$\sigma$ and 0.153 mag respectively. 

%I changed the number of sig figs on the scatter reports above and below, otherwise all 3 are 0.15 and it was kind of weird when we claimed the last was the smallest! -Cole 

Finally, we fit a 2-D linear correlation in $A_V$ and $\delta$ to the Hubble Residuals simultaneously and find 
$\Delta\mu=-0.075A_V+0.197\delta+0.184$. This provides the smallest scatter {(0.150 mag)} amongst the three correlations tested. Step size is also dramatically reduced, down to $\sim 2.1\sigma$. 

{\citet{salim2020dust} already noted the stellar mass dependence of the attenuation law slope (i.e., similarly for $\delta$ and $R_V$) is driven by the observed relation between slope and $A_V$. In other words, the stellar mass correlation with dust parameters is driven by $A_V$. We interpret our findings in light of \citet{Salim18,salim2020dust}: the Hubble residuals dependence on $A_V$ is most effective in reducing the mass step because $A_V$ drives the correlation between dust law and stellar mass.} Evidently, there is a connection between the dust parameters of the host galaxy and the mass step. {These results offer a compelling physical mechanism for understanding the mass step and also show that host galaxy dust properties have the potential to be used to reduce the scatter in the Hubble diagram.}

\subsection{Challenges with measurements of dust parameters from broadband photometry}

It is well known that there exist degeneracies between several galaxy parameters, especially as measured from optical/NIR broadband photometry (e.g., \citealt{Qin_2022}). Because of the way the dust parameters explored here are defined, it is obvious that there will be a certain degree of degeneracy between $\delta$ and $A_V$. SFR is another parameter that is degenerate with the dust parameter, as a red and quenched galaxy with little dust content may look very similar to a star forming galaxy with a steep dust attenuation law in optical/NIR photometry. A galaxy sample such as the one presented by \citet{Salim18,salim2020dust} has exquisite photometry throughout a wide range of filters, which makes it possible to break those parameter degeneracies, especially with the inclusion of longer wavelength, IR bands. Their robust analysis shows that the correlations found between $\delta$ and $A_V$ are not an artifact. Our results are widely in agreement with their findings, despite the very different datasets and techniques used, which supports the conclusion that this work is also not retrieving some spurious relations. To further test this conclusion, we test our method on a state-of-the-art panchromatic extragalactic catalog, the COSMOS catalog from \citet{Weaver_2022}.

Our test consists in mapping our SN hosts onto similar galaxies in the COSMOS catalog, and then comparing the derived galaxy properties of the two samples. By doing so, we hope to understand whether similar trends are found at the population level when a similar galaxy sample, with significantly improved photometric data including IR bands, is analyzed. First, for each SN host, we find the top 2 nearest neighbours in the DES COSMOS data based on their magnitude, in redshift bins. We then find the COSMOS data from \citet{Weaver_2022} that matches the DES COSMOS galaxies, and run BAGPIPES using all of the available COSMOS data. We assume the COSMOS photometric redshifts as priors in the BAGPIPES run. Because the COSMOS field is only 2 deg$^2$, we do not expect to find appropriate neighbour matches for all galaxies, especially those at the lowest redshifts, so some differences between the SN hosts and the COSMOS populations are expected. We are, however, able to recover with the COSMOS neighbours population: 1) two significantly different distributions for low versus high mass galaxies for the dust parameters shown in Fig. \ref{fig:5}; 2) similar trends for $A_V$ and $\delta$ versus mass as those in Section \ref{sec:results}; 3) 
Hubble residuals steps and trends similar to those in \ref{fig:5},
although at a lower significance. We therefore conclude that, although there may be degeneracies amongst dust and star formation parameters due to the limited photometric bands available, the trends are confirmed when using data from one of the richest photometric galaxy catalogs available, and are likely not spurious. In other words, although the dust parameters may not be well constrained for each single galaxy, as indicated by the large uncertainties in e.g., Fig. \ref{fig:5}, the overall trends found for the high versus low mass galaxies and for the step function are not expected to be artifacts.

\begin{table*}
\centering
\begin{tabular}{llll}
\hline
\hline
Dust Prediction from SN Ia & Reference & Global Galaxy Dust Finding & Reference \\ % %& 
\hline

$R_V$ ranges from $\sim$1-6 & BS21 & $R_V$ ranges from $\sim$1-6  &   Fig. \ref{fig:5} (this work) and  \\
& &  &   \cite{salim2020dust} \\ \\

High mass galaxies mean $R_V$ is $\sim$0.5 & BS21 & High mass galaxies mean $R_V$ is $\sim0.7$  & Fig. \ref{fig:5} (this work) and\\
lower than low mass galaxies  &  &  lower than low mass galaxies; high &  \cite{salim2020dust}\\ 
 &  &  mass sample has two $R_V$ populations & \\ \\

The mass step  & BS21 & Correcting for galaxy dust properties & Fig. \ref{fig:rm} and \ref{dustfits} (this work) \\
is driven by dust differences&  & reduces mass step significance to $2\sigma$   &  \\ \\

Host color - SN Ia Hubble Residual step is  & \citet{Roman_2018} & Host color is strongly correlated  & Fig. \ref{fig:rvhostcolor} (this work)\\ 
as large or larger than mass step & & with host $R_V$ & \\ \\

The SN Ia color-luminosity relation & BS21 & Red-sequence galaxies have very low $R_V$, & Fig. \ref{fig:rm} (this work) and  \\
 ($\beta$) is largely driven by $R_V$ &  & and measurement of $\beta$ for same sample  &  and \cite{Chenetal} \\ 
&&is very low&\\ \\

SNe Ia in red-sequence & \cite{Chenetal} &   Red sequence galaxies have smaller &  Fig. \ref{fig:rm} (this work) \\
 galaxies have less HR scatter  & & range of $R_V$ than other galaxies &   \\
 than SNe Ia in other galaxies&& & \\ \\

\hline
\hline
\end{tabular}\caption{Comparison of the findings from galaxies studies on dust, to relevant SNIa predictions. HR stands for `Hubble Residual'. }\label{table:comparing}
\end{table*}

\section{Discussion and Conclusions}\label{sec:conc}

In this work, we have presented a detailed study of DES SN hosts, posing particular attention to the estimation of dust parameters and to correlations of dust parameters with SN Ia Hubble diagram residuals. {While some host galaxy properties, such as dust, star formation histories, and stellar mass, have been recently explored with simulations or control samples,} this is the first work { exploring the impact of dust estimates from host photometry alone on the SN Ia mass step.} 
We detect significant correlations between global host galaxy dust parameters and Hubble residuals for the first time. We show how these correlations are able to account for the mass step to a large extent, and how host galaxy dust parameters can be used to reduce the scatter in the SNe Ia Hubble diagram.

{ We show in Table \ref{table:comparing} the predictions from SNe Ia and compare them with the findings for host dust properties; all of which are consistent with the SN Ia color-luminosity relation, SN Ia intrinsic scatter, and SN Ia `mass step' being driven by host-dust.} The following points summarize our findings:
\begin{itemize}
    \item There is significant ($\gtrsim 4\sigma$) step-like behaviour in the Hubble residuals as a function of dust parameters $A_V$ and $\delta$ (and consequently, $R_V$). These step-like relations are driven by the red SNe (with color parameter $c > -0.025$), while they wash out for the bluer SNe. This finding, together with the fact that the step is as strong or stronger than the mass step, could be indicating that these relations are due to the presence of dust, rather than being a simple consequence of the underlying correlation between host galaxy mass and dust.  
    \item There exists a linear correlation, significant at $3\sigma$, between the Hubble residual and the dust parameter $\delta$ when one removes SNe where dust attenuation is negligible, i.e. the bluer SNe with color parameter $c < -0.025$.
    \item Correlations of host galaxy dust parameters with the Hubble residuals can account for the size of the mass step { and simultaneously reduce scatter in red SNe}, indicating that the origin of the mass step could be in the host galaxy dust. {Due to the small sample size, this study is unable to confirm or rule out the presence of other mechanisms that could be partially responsible for the mass step, along with dust. Future analyses with larger data sets should be able to provide a more quantitative statement on the extent to which the mass step can be explained by dust alone.}
    \item Host galaxy dust parameters can be used to reduce the scatter in the Hubble diagram by up to $13\%$, thus potentially improving SN cosmology results.
    \item We find evidence for two different populations of $\delta$ (or equivalently, $R_V$) when separating the sample into low and high mass galaxies, following a typical mass step split at $10^{10}~M_\odot$. These two populations show a similar behaviour to what  \citet{brout2020dust} predicted would be observed if dust were to explain the observed mass step: lower mass galaxies tend to have a larger $\delta$ (or equivalently, $R_V$).
    \item These results for $\delta$ are similar to what is found in a generic galaxy sample. Moreover, SN Ia hosts show a strong correlation between the dust attenuation in $V$ band $A_V$ and stellar mass, which is also consistent with a generic galaxy population. 
    \item We find that the $A_V$ derived from the integrated galaxy photometry is not strongly correlated with the amount of dust along the line of sight to the SNe (SALT2 color). It is important to note that host-galaxy attenuation and its wavelength dependence ($R_V$) in this work have been derived from the non-local galaxy photometry, yet we compare our results with the dust extinction and properties derived from line of sight photometry to the SNe. Our findings agree with the general expectation that while the amount of dust can vary greatly across a galaxy, the observed SN color is highly dependent on its location and line of sight within the galaxy. This is also consistent with findings of sibling SNe in the same galaxy having large variation in SN colors \citep{scolnicsiblings,Biswas_2021,scolnicpanplus}. However, we do find a significant correlation between the $R_V$ (or $\beta$) found from the SNe and that from the attenuation of the whole galaxy. We interpret this as a consequence of the fact that the size and properties of the dust grains themselves, which are expected to drive $R_V$ \citep{Salim18}, are similar across the galaxy and any single location or line of sight to the SN is a representative probe.
    \item SN host galaxies' $\delta$ and $R_V$ span a wide range of values: while lower mass galaxies tend to have $-0.5<\delta< 0.1$ or $2<R_V<4$ (68\% CI of the distribution), a significant (of the order of $10\%$) fraction of galaxies with mass $>10^{10}~M_\odot$ reach below $\delta\sim -0.9$ or $R_V<2$. Our findings suggest that assuming a single value or a very narrow distribution for $R_V$, as often done is SN cosmology, may not be accurate.
%%%Removed for final reading
    %\item Models for reddening $E(B-V)$ distributions can be improved by taking into account the host galaxy's reddening. We present a best fit for our $E(B-V)$ distribution that follows a log-normal function.

\end{itemize}

A wider area, wider wavelength coverage (including more infrared bands than VISTA), { and deeper} photometric data set than what used here is under preparation for the final DES deep-field catalogs that will be used for the final DES cosmology analyses. This catalog will cover 3 times as many SN Ia hosts as those used in this work, and it will enable more precise dust parameter estimation for each galaxy. As such, this catalog will be a valuable dataset to replicate this analysis and confirm with more precision and confidence the results of this work. We will also be able to extend the analyses to use the Hubble residuals for the photometric sample of \cite{moller22}, including more than 1000 SNe, as opposed to the 81 spectroscopically classified SNe used in the last part of this work. Precision spectroscopy of host galaxies in the optical to infrared, and redder bands (at $\gtrsim 8~\mu$m in rest frame), will also be valuable to derive more precise dust properties constraints in SN hosts and therefore strengthen the argument that the observed mass step is a consequence of dust properties in SN host galaxies. Measurements of this kind will also be useful to improve measurements of cosmological parameters from the SN Ia Hubble diagram, since we have shown that it is possible to use dust parameters information to improve the standardization of SN Ia. We conclude that a better understanding of the origin of the mass step and its correlation with dust may be crucial to obtain more precise and accurate cosmological constraints from SN cosmology.

\section*{Acknowledgments}
CM was supported by the University of Chicago Astronomy \& Astrophysics Department for an internship at Fermilab. We thank Adam Carnall for making BAGPIPES public and for their help with the code. We thank Samir Salim and Charlie Conroy for enlightning discussion. We would also like to thank Bang Wong for a color scheme for our plots that is optimized for readability for the colorblind \citep{colorblindness}. 

Antonella Palmese acknowledges support for this work was provided by NASA through the NASA Hubble Fellowship grant HST-HF2-51488.001-A awarded by the Space Telescope Science Institute, which is operated by Association of Universities for Research in Astronomy, Inc., for NASA, under contract NAS5-26555.
Dillon Brout acknowledges support for this work was provided by NASA through the NASA Hubble Fellowship grant HST-HF2-51430.001 awarded by the Space Telescope Science Institute, which is operated by Association of Universities for Research in Astronomy, Inc., for NASA, under contract NAS5-26555. Simulations, light-curve fitting, BBC, and cosmology pipeline managed by \texttt{PIPPIN} \citep{Pippin}.

Funding for the DES Projects has been provided by the U.S. Department of Energy, the U.S. National Science Foundation, the Ministry of Science and Education of Spain, 
the Science and Technology Facilities Council of the United Kingdom, the Higher Education Funding Council for England, the National Center for Supercomputing 
Applications at the University of Illinois at Urbana-Champaign, the Kavli Institute of Cosmological Physics at the University of Chicago, 
the Center for Cosmology and Astro-Particle Physics at the Ohio State University,
the Mitchell Institute for Fundamental Physics and Astronomy at Texas A\&M University, Financiadora de Estudos e Projetos, 
Funda{\c c}{\~a}o Carlos Chagas Filho de Amparo {\`a} Pesquisa do Estado do Rio de Janeiro, Conselho Nacional de Desenvolvimento Cient{\'i}fico e Tecnol{\'o}gico and 
the Minist{\'e}rio da Ci{\^e}ncia, Tecnologia e Inova{\c c}{\~a}o, the Deutsche Forschungsgemeinschaft and the Collaborating Institutions in the Dark Energy Survey. 

The Collaborating Institutions are Argonne National Laboratory, the University of California at Santa Cruz, the University of Cambridge, Centro de Investigaciones Energ{\'e}ticas, 
Medioambientales y Tecnol{\'o}gicas-Madrid, the University of Chicago, University College London, the DES-Brazil Consortium, the University of Edinburgh, 
the Eidgen{\"o}ssische Technische Hochschule (ETH) Z{\"u}rich, 
Fermi National Accelerator Laboratory, the University of Illinois at Urbana-Champaign, the Institut de Ci{\`e}ncies de l'Espai (IEEC/CSIC), 
the Institut de F{\'i}sica d'Altes Energies, Lawrence Berkeley National Laboratory, the Ludwig-Maximilians Universit{\"a}t M{\"u}nchen and the associated Excellence Cluster Universe, 
the University of Michigan, NFS's NOIRLab, the University of Nottingham, The Ohio State University, the University of Pennsylvania, the University of Portsmouth, 
SLAC National Accelerator Laboratory, Stanford University, the University of Sussex, Texas A\&M University, and the OzDES Membership Consortium.

Based in part on observations at Cerro Tololo Inter-American Observatory at NSF's NOIRLab (NOIRLab Prop. ID 2012B-0001; PI: J. Frieman), which is managed by the Association of Universities for Research in Astronomy (AURA) under a cooperative agreement with the National Science Foundation.

The DES data management system is supported by the National Science Foundation under Grant Numbers AST-1138766 and AST-1536171.
The DES participants from Spanish institutions are partially supported by MICINN under grants ESP2017-89838, PGC2018-094773, PGC2018-102021, SEV-2016-0588, SEV-2016-0597, and MDM-2015-0509, some of which include ERDF funds from the European Union. IFAE is partially funded by the CERCA program of the Generalitat de Catalunya.
Research leading to these results has received funding from the European Research
Council under the European Union's Seventh Framework Program (FP7/2007-2013) including ERC grant agreements 240672, 291329, and 306478.
We  acknowledge support from the Brazilian Instituto Nacional de Ci\^encia
e Tecnologia (INCT) do e-Universo (CNPq grant 465376/2014-2).

This manuscript has been authored by Fermi Research Alliance, LLC under Contract No. DE-AC02-07CH11359 with the U.S. Department of Energy, Office of Science, Office of High Energy Physics.

\appendix
\section{Comparison of hosts photometry}\label{sec:appendix}
%\appendixpage
%\noappendicestocpagenum

\begin{figure*}
    \includegraphics[width=.6\textwidth]{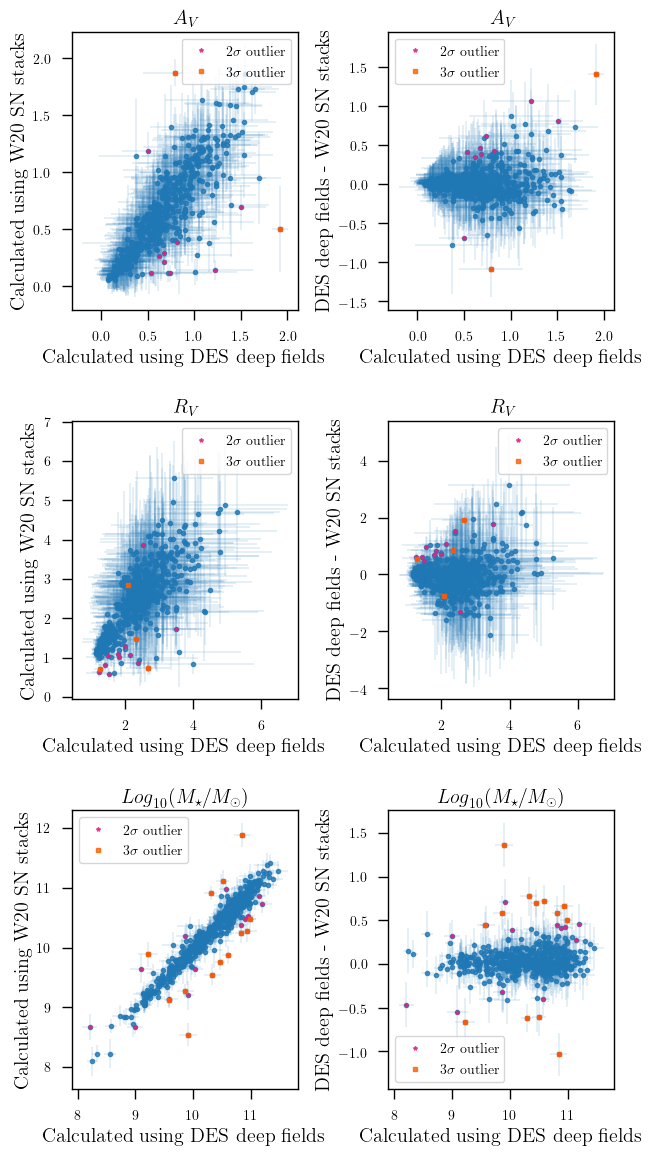}
    \caption{Comparison of host galaxy properties computed with two different sets of photometric data: that of the DES deep fields of \citet*{hartley2020dark} (shown on the $x$ axis of each panel), and that from the catalog of \citet{wiseman20} (shown on the $y$ axis of each panel on the left hand side). The right-hand side shows, on th $y$-axis, the residuals given by the difference between the two estimations of the same parameter. We show results for the dust parameters $A_V$ and $R_V$, and the stellar mass, as these are the relevant quantities for this work. The magenta and orange points represent the 2 and 3 $\sigma$ outliers, respectively. }
    \label{filt_comparison}
\end{figure*}

In this section we compare the host galaxy properties computed with two different sets of photometric data, as shown in Figure \ref{filt_comparison}: that of the DES deep fields of \citet*{hartley2020dark} (on the $x$ axis of each panel), and that from the catalog of \citet{wiseman20} (on the $y$ axis of each panel). We remove objects that are brighter in the deep fields catalog (i.e. likely contaminated by the SN light) by more than twice the standard deviation of the residual magnitude distribution from the comparison of the two photometric datasets in any band. Switching between photometric data does not introduce a large amount of variation: only 1, 2, and 5\% of galaxies had their $A_V, R_V$ or stellar mass respectively changed by more than 2$\sigma$ (magenta and orange points in Figure \ref{filt_comparison}). We choose the SN Stack data in our fiducial dataset throughout the paper as it removes contamination due to the supernova that contaminates light from the host galaxy. While it is not obvious which threshold on the outlier fraction should be used in order to consider the change problematic for this work, we have made the entire analysis in this work using the photometry from the DES deep fields, and did not find any significant change in our conclusions. Moreover, the $\chi^2$ cut we apply on the BAGPIPES outputs results in a reduction in the number of galaxies in the deepfields case compared to the W20 catalog because of the contaminated photometry, so that the W20 also provides better fits to the galaxies SEDs.

\section{Stellar Mass - Dust Parameter Relations for an Exponential Star Formation history}\label{sec:appendixB}

\begin{figure*}
    \includegraphics[width=1\textwidth]{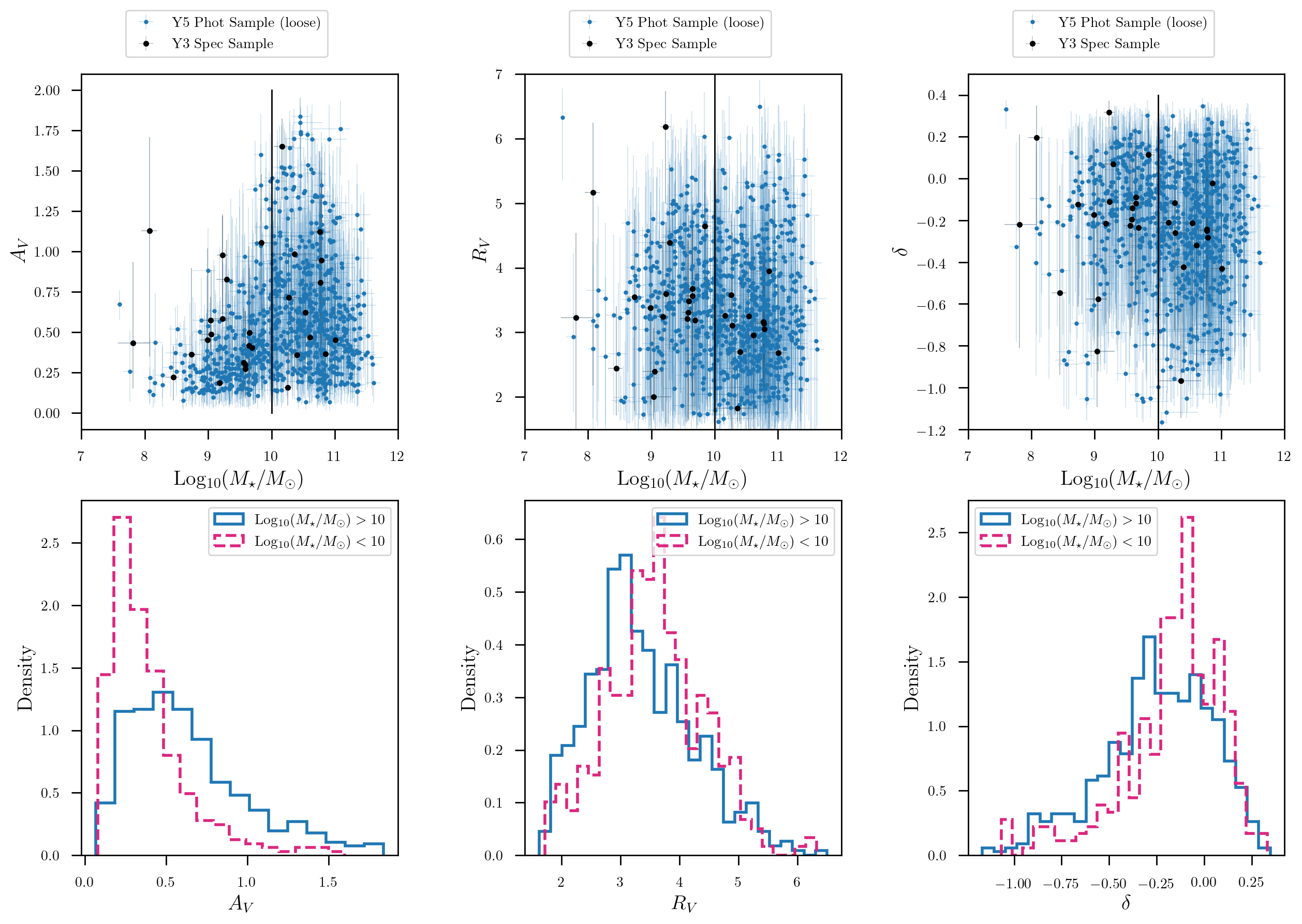}
    \caption{Same as Figure \ref{fig:5}, but with parameters derived assuming an exponential SFH. Top: Dust parameters $A_V$ and $R_V$ as a function of stellar mass, derived using an exponential SFH. Bottom: Distribution of $A_V$ and $R_V$ values above and below the $10^{10} M_{\odot}$ split, the location of the observed mass step.}
    \label{fig:9}
\end{figure*}

While in Figure \ref{fig:5} we assume a log-normal SFH, Figure \ref{fig:9} contains similar plots for an exponential SFH with a prior of $\tau<2$ Gyr. In this figure, it is clear that a different assumption in the SFH makes the correlation between the dust parameters and the stellar mass less obvious, at least when splitting the sample at $10^{10}M_\odot$. We interpret this result as a consequence of the fact that there exists a degeneracy between the dust, the SFH, and the stellar mass of galaxies when their SEDs are fit using the available photometric bands. We note that, however, as soon as the prior on $\tau$ is relaxed to a more realistic one (e.g. $\tau<10$ Gyr), the $A_V/\delta$ versus mass plots resemble more closely our findings in Figure \ref{fig:5}. We conclude that that the main findings of this work are not largely affected by the exact choice of the SFH parameterisation, as long as the prior choices are realistic. We show here the effect of a more extreme and unrealistic SFH choice (although used in the literature, e.g. S20) on the recovered parameters, when only more peaked SFHs are considered (because of the low $\tau$ values allowed).

\section{Hubble residuals and dust parameters with different host galaxy SED fitting priors}
\label{appc}

\begin{figure*}
    \centering
    \includegraphics[width=.85\linewidth]{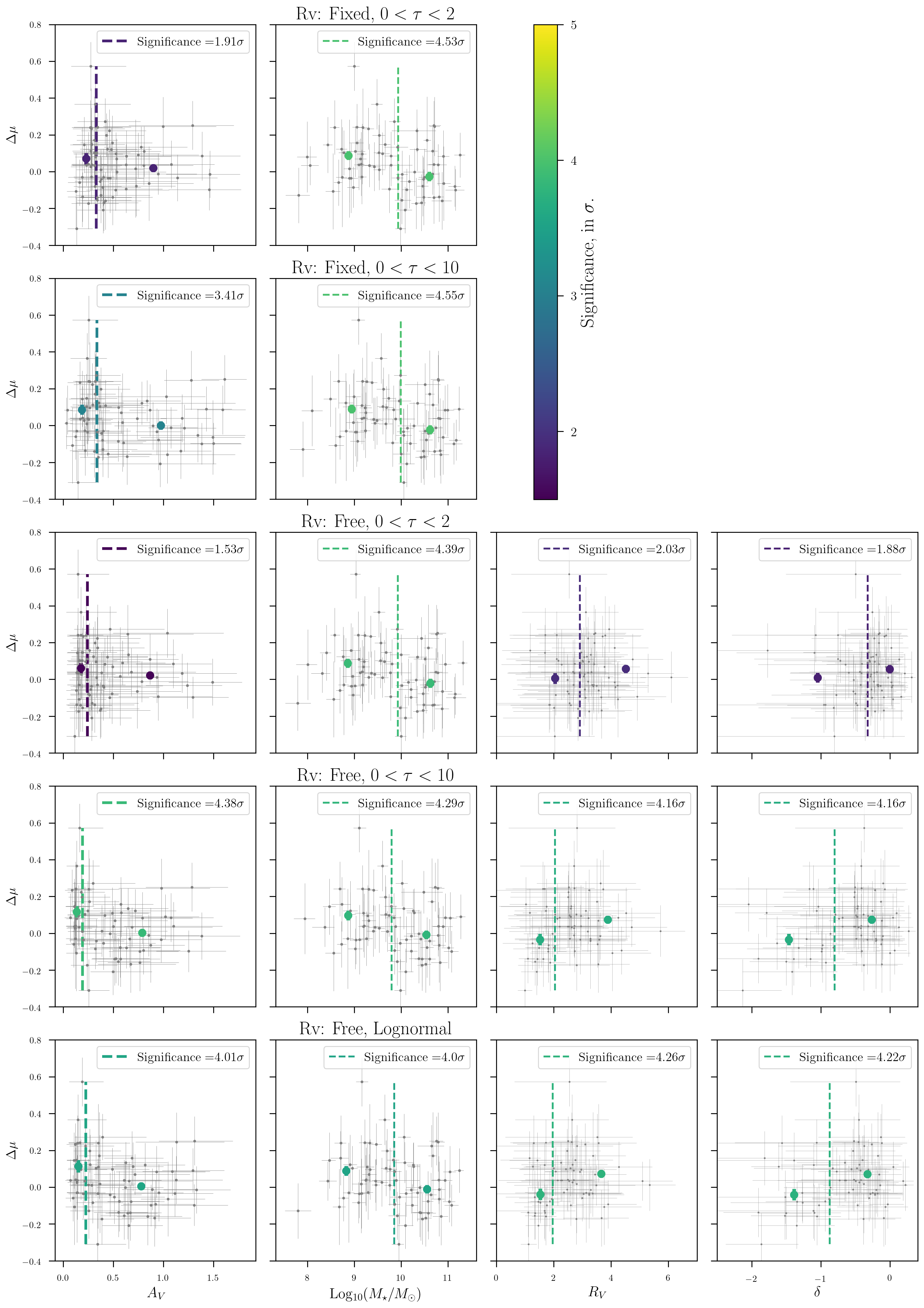}
    \caption{Step-like behaviour in Hubble residuals as a function of $A_V$, stellar mass, $R_V$ and $\delta$. These are calculated using five SFHs, four of which are exponential with $0<\tau<2$ or $0<\tau<10$ Gyr with $R_V$ free or fixed, the final being log-normal with $R_V$ left free. We then scan over each parameter to {search for a significant step in the Hubble Residuals, i.e. a certain point where the mean of the Hubble residuals on either side of said point is significantly different}. The most significant step found is displayed in each panel, along with its 3$\sigma$ errorbars, and is coloured according to its significance. } 
    \label{appc_fig}
\end{figure*}

In this Appendix we study the correlations between Hubble residuals and various dust parameters assuming different SFHs and dust parameters priors. 
Similarly to Figure \ref{fig:12}, Figure \ref{appc_fig} shows the step functions for host galaxy parameters computed with different assumptions: an exponential SFH with $0<\tau<2$ Gyr, with $R_V$ both free and fixed, an exponential SFH with $0<\tau<10$ Gyr, with $R_V$ free and fixed, and a log-normal SFH with $R_V$ left free. The form of the SFH has a noticeable effect on the significance of the steps in the Hubble Residuals. We find a significant step with respect to mass in every SFH, but for those where $\tau$ is restricted to [0,2] Gyr there is no significant step in the dust parameters. In every other case where there is no such restriction on $\tau$ there is a significant step in all dust parameters. This is important to note for the DES cosmology, as the $0 < \tau < 2$ Gyr prior is the one used in S20, where the mass step is described in detail as resulting from the DES 3Yr data. 

\begin{figure*}
    \centering
    \includegraphics[width=.85\linewidth]{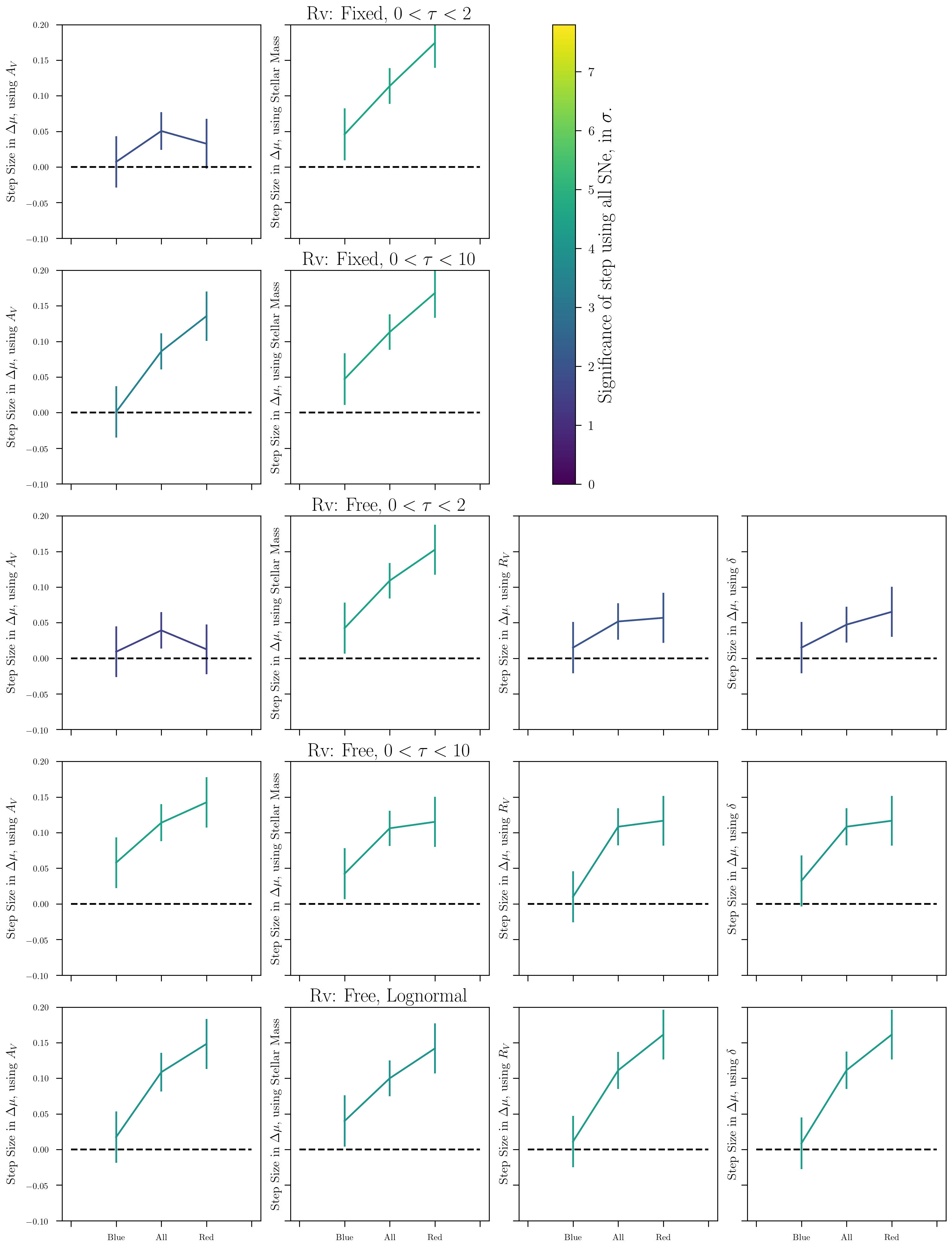}
    \caption{Step size in the Hubble Residuals as we move from what we expect to be a less dusty to a more dusty sample. The ``red'' and ``blue'' samples refer to red and blue SNe, having a colour above or below -0.025. The ``All'' sample refers to what was shown in Figure \ref{fig:12}. 
    As in Figure \ref{fig:12}, the steps are coloured according to their significance where significance was determined by the step calculated from the SNe and host properties in each subsample. }
    \label{appc_2}
\end{figure*}

We also show the equivalent of the bottom panels in Figure \ref{fig:12} but for multiple SFH and parameter prior combinations in Figure \ref{appc_2}. In this case, we show the step size, with its uncertainty, after splitting the sample into hosts of red (with color parameter $c>-0.025$, presumably in a dustier environment) and blue (with color parameter $c<-0.025$, presumably in a less dusty environment) SNe. There are a few important conclusions to be drawn from these plots. 
Firstly, all the plots 
that were considered significant in Figure \ref{appc_fig} have sharp upward trends in step size as we move to supposedly ``dustier'' SN environments. This is further evidence that the mass step has some dependence on dust parameters.
Secondly, the cases that were not deemed significant do not experience this trend. We define as significant a step where the mean values on each point are at least $3\sigma$ away from each other. The two non-significant cases in $A_V$ (the leftmost column) lack the increase in step when going from a less dusty to a dustier environment on average, while the two non-significant cases in the two rightmost columns (they are actually the same, since they are $\delta$ and $R_V$ for the same case) increase very slightly but stay non-significant in all three samples. 
Since there is no theoretical or empirical reason to restrict $\tau$ to $<2$ Gyr, we consider the results assuming the log-normal SFH (or similarly, the exponential SFH with $0<\tau<10$ Gyr), as our fiducial results, also considering how the change in SFH parametrisation does not appear to have a noticeable effect on the step size or significance.

\section*{Affiliations}
$^{1}$ Department of Astronomy and Astrophysics, University of Chicago, Chicago, IL 60637, USA\\
$^{2}$ Fermi National Accelerator Laboratory, P. O. Box 500, Batavia, IL 60510, USA\\
$^{3}$ Kavli Institute for Cosmological Physics, University of Chicago, Chicago, IL 60637, USA\\
$^{4}$ Department of Astronomy, University of California, Berkeley,  501 Campbell Hall, Berkeley, CA 94720, USA\\
$^{5}$ Center for Astrophysics $\vert$ Harvard \& Smithsonian, 60 Garden Street, Cambridge, MA 02138, USA\\
$^{6}$ Department of Physics, Duke University Durham, NC 27708, USA\\
$^{7}$ Institute of Cosmology and Gravitation, University of Portsmouth, Portsmouth, PO1 3FX, UK\\
$^{8}$ Institut d'Estudis Espacials de Catalunya (IEEC), 08034 Barcelona, Spain\\
$^{9}$ Institute of Space Sciences (ICE, CSIC),  Campus UAB, Carrer de Can Magrans, s/n,  08193 Barcelona, Spain\\
$^{10}$ Department of Astronomy, University of Geneva, ch. d'\'Ecogia 16, CH-1290 Versoix, Switzerland\\
$^{11}$ School of Mathematics and Physics, University of Queensland,  Brisbane, QLD 4072, Australia\\
$^{12}$ Centre for Astrophysics \& Supercomputing, Swinburne University of Technology, Victoria 3122, Australia\\
$^{13}$ Centre for Gravitational Astrophysics, College of Science, The Australian National University, ACT 2601, Australia\\
$^{14}$ The Research School of Astronomy and Astrophysics, Australian National University, ACT 2601, Australia\\
$^{15}$ Lawrence Berkeley National Laboratory, 1 Cyclotron Road, Berkeley, CA 94720, USA\\
$^{16}$ School of Physics and Astronomy, University of Southampton,  Southampton, SO17 1BJ, UK\\
$^{17}$ Centro de Investigaciones Energ\'eticas, Medioambientales y Tecnol\'ogicas (CIEMAT), Madrid, Spain\\
$^{18}$ INAF-Osservatorio Astronomico di Trieste, via G. B. Tiepolo 11, I-34143 Trieste, Italy\\
$^{19}$ Sydney Institute for Astronomy, School of Physics, A28, The University of Sydney, NSW 2006, Australia\\
$^{20}$ Laborat\'orio Interinstitucional de e-Astronomia - LIneA, Rua Gal. Jos\'e Cristino 77, Rio de Janeiro, RJ - 20921-400, Brazil\\
$^{21}$ Department of Physics, University of Michigan, Ann Arbor, MI 48109, USA\\
$^{22}$ CNRS, UMR 7095, Institut d'Astrophysique de Paris, F-75014, Paris, France\\
$^{23}$ Sorbonne Universit\'es, UPMC Univ Paris 06, UMR 7095, Institut d'Astrophysique de Paris, F-75014, Paris, France\\
$^{24}$ University Observatory, Faculty of Physics, Ludwig-Maximilians-Universit\"at, Scheinerstr. 1, 81679 Munich, Germany\\
$^{25}$ Department of Physics \& Astronomy, University College London, Gower Street, London, WC1E 6BT, UK\\
$^{26}$ Kavli Institute for Particle Astrophysics \& Cosmology, P. O. Box 2450, Stanford University, Stanford, CA 94305, USA\\
$^{27}$ SLAC National Accelerator Laboratory, Menlo Park, CA 94025, USA\\
$^{28}$ Institut de F\'{\i}sica d'Altes Energies (IFAE), The Barcelona Institute of Science and Technology, Campus UAB, 08193 Bellaterra (Barcelona) Spain\\
$^{29}$ Center for Astrophysical Surveys, National Center for Supercomputing Applications, 1205 West Clark St., Urbana, IL 61801, USA\\
$^{30}$ Department of Astronomy, University of Illinois at Urbana-Champaign, 1002 W. Green Street, Urbana, IL 61801, USA\\
$^{31}$ Astronomy Unit, Department of Physics, University of Trieste, via Tiepolo 11, I-34131 Trieste, Italy\\
$^{32}$ Institute for Fundamental Physics of the Universe, Via Beirut 2, 34014 Trieste, Italy\\
$^{33}$ Department of Physics, IIT Hyderabad, Kandi, Telangana 502285, India\\
$^{34}$ Jet Propulsion Laboratory, California Institute of Technology, 4800 Oak Grove Dr., Pasadena, CA 91109, USA\\
$^{35}$ Institute of Theoretical Astrophysics, University of Oslo. P.O. Box 1029 Blindern, NO-0315 Oslo, Norway\\
$^{36}$ Instituto de Fisica Teorica UAM/CSIC, Universidad Autonoma de Madrid, 28049 Madrid, Spain\\
$^{37}$ Department of Physics and Astronomy, University of Pennsylvania, Philadelphia, PA 19104, USA\\
$^{38}$ Observat\'orio Nacional, Rua Gal. Jos\'e Cristino 77, Rio de Janeiro, RJ - 20921-400, Brazil\\
$^{39}$ Santa Cruz Institute for Particle Physics, Santa Cruz, CA 95064, USA\\
$^{40}$ Center for Cosmology and Astro-Particle Physics, The Ohio State University, Columbus, OH 43210, USA\\
$^{41}$ Department of Physics, The Ohio State University, Columbus, OH 43210, USA\\
$^{42}$ Australian Astronomical Optics, Macquarie University, North Ryde, NSW 2113, Australia\\
$^{43}$ Lowell Observatory, 1400 Mars Hill Rd, Flagstaff, AZ 86001, USA\\
$^{44}$ George P. and Cynthia Woods Mitchell Institute for Fundamental Physics and Astronomy, and Department of Physics and Astronomy, Texas A\&M University, College Station, TX 77843,  USA\\
$^{45}$ Instituci\'o Catalana de Recerca i Estudis Avan\c{c}ats, E-08010 Barcelona, Spain\\
$^{46}$ Physics Department, 2320 Chamberlin Hall, University of Wisconsin-Madison, 1150 University Avenue Madison, WI  53706-1390\\
$^{47}$ Institute of Astronomy, University of Cambridge, Madingley Road, Cambridge CB3 0HA, UK\\
$^{48}$ Hamburger Sternwarte, Universit\"{a}t Hamburg, Gojenbergsweg 112, 21029 Hamburg, Germany\\
$^{49}$ Computer Science and Mathematics Division, Oak Ridge National Laboratory, Oak Ridge, TN 37831\\
$^{50}$ Excellence Cluster Origins, Boltzmannstr.\ 2, 85748 Garching, Germany\\
$^{51}$ Max Planck Institute for Extraterrestrial Physics, Giessenbachstrasse, 85748 Garching, Germany\\
$^{52}$ Universit\"ats-Sternwarte, Fakult\"at f\"ur Physik, Ludwig-Maximilians Universit\"at M\"unchen, Scheinerstr. 1, 81679 M\"unchen, Germany\\

\bibliographystyle{mnras}
\bibliography{references}
\end{document}